\documentclass[preprint]{aastex}

\usepackage{float}
\usepackage{bm}
\usepackage{color}
\usepackage{amsmath}
\usepackage{amssymb}
\usepackage{graphicx}
\usepackage{natbib}
\usepackage{framed}

\newcommand{\lamp}{LAMP}
\newcommand{\nw}{nW m$^{-2}$ sr$^{-1}$}
\newcommand{\eps}{e$^{-}$s$^{-1}$}
\newcommand{\oh}{OH}
\newcommand{\lamplambda}{$1191.3 \,$nm}
\newcommand{\mkk}{$M_{\ell \ell^{\prime}}$}
\newcommand{\magas}{mag$_{\rm AB}$ arcsec$^{-2}$}

\slugcomment{Submitted to PASP July 16, 2015.}

\shorttitle{Stability of Airglow at \lamplambda}
\shortauthors{Nguyen et al.}

\begin{document}

\title{Spatial and Temporal Stability of Airglow Measured in
  the Meinel Band Window at \lamplambda}

\author {Hien T.~Nguyen\altaffilmark{1,2 \ast},
  Michael Zemcov\altaffilmark{2,1}, John Battle\altaffilmark{2},
  James J.~Bock\altaffilmark{2,1}, Viktor Hristov\altaffilmark{2},
  Philip Korngut\altaffilmark{1,2}, and Andrew Meek\altaffilmark{2}}
\affil{$^{1}$Jet Propulsion Laboratory (JPL), National Aeronautics and Space
  Administration (NASA), Pasadena, CA 91109, USA}
\affil{$^{2}$Department of Physics, Mathematics and Astronomy, California
Institute of Technology, Pasadena, CA 91125, USA}
\email{$^{\ast}$htnguyen@jpl.nasa.gov}

\begin{abstract}
  We report on the temporal and spatial fluctuations in the
  atmospheric brightness in the narrow band between Meinel emission lines at
  \lamplambda\ using a $\lambda / \Delta \lambda=320$ 
  near-infrared instrument.  We present the instrument design and
  implementation, followed by a detailed analysis of data taken over
  the course of a night from Table Mountain Observatory.  The absolute
  sky brightness at this wavelength is found to be $5330 \pm 30
  \,$\nw, consistent with previous measurements of the inter-band
  airglow at these wavelengths.  This amplitude is larger than simple
  models of the continuum component of the airglow emission at these
  wavelengths, confirming that an extra emissive or scattering
  component is required to explain the observations.  We perform a
  detailed investigation of the noise properties of the data and find
  no evidence for a noise component associated with temporal
  instability in the inter-line continuum.  This result demonstrates
  that in several hours of $\sim 100 \,$s integrations the noise
  performance of the instrument does not appear to significantly
  degrade from expectations, giving a proof of concept that near-IR
  line intensity mapping may be feasible from ground-based sites.
\end{abstract}


\section{Introduction}

Recent results (e.g.~\citealt{Kashlinsky2005},
\citealt{Matsumoto2011}, \citealt{Zemcov2014}) suggest there are
large-angular scale fluctuations in the near-infrared (IR)
extragalactic background light (EBL) larger than models of galaxy
clustering predict \citep{Helgason2012}.  This component appears to
reach a maximum brightness between $1$ and $2 \, \mu$m, wavelengths at
which terrestrial airglow is very bright \citep{Leinert1998}.  As a
result, the aforementioned measurements of the large-angular emission
have only been performed from space-based platforms in broad
photometric bands.  Due to the complexity of space-based measurements,
such observations are expensive.  As a result, the ability to make
ground-based observations of this component would significantly
expedite progress.

Airglow in the near-IR prohibits measurement of large-angular scale
structure in broad bands from the ground.  In the range
$800 < \lambda < 2000 \,$nm, the predominant mechanism responsible for
airglow is emission from reactions between O$_{3}$ and H in the upper
atmosphere \citep{LeTexier1987}, called Meinel emission
\citep{Meinel1950}.  Because of the properties of oxygen mixing in the
atmosphere, airglow is produced in a discrete layer between 75 and 100
km \citep{vonSavigny2012}, so observations made from different
ground-based sites should see similar brightness.  The Meinel spectrum
exhibits a large number of narrow lines, which averaged over wide
bands lead to backgrounds of $\sim 15 \,$\magas.  However, in
$\lambda / \Delta \lambda \sim 100$ windows between the lines the sky
is stable and approaches a surface brightness of $\sim 20 \,$\magas\
\citep{Sullivan2012}.  If the \oh\ line positions are sufficiently
stable, it should be possible to perform imaging measurements in the
continuum windows between them.

\begin{figure*}[p]
   \centering
   \resizebox{0.7\textheight}{!}{\includegraphics{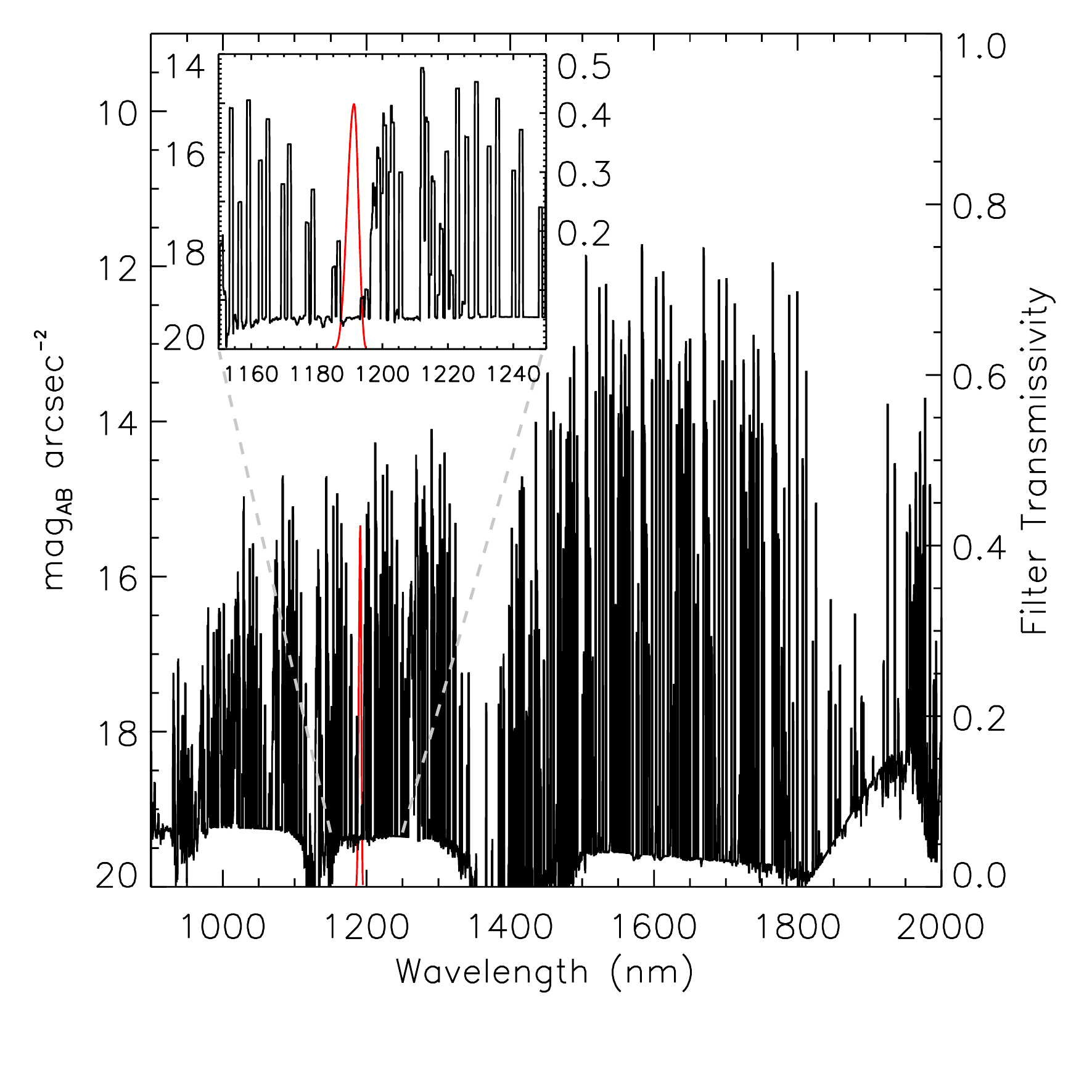}}
   \caption{The predicted surface brightness of the atmosphere between
     $0.9$ and $2.0 \, \mu$m.  The atmospheric emission spectrum is
     derived from the Gemini Observatory sky background model (black
     line).  This model is generated using the sky transmission files
     generated by ATRAN \citep{Lord1992} scaled to a $273 \,$K
     blackbody. We assume $5 \,$mm of precipitable water vapor in this
     calculation, which is consistent with the typical value above
     Table Mountain during clear weather \citep{Leblanc2011}.  An OH
     emission spectrum, a set of O$_{2}$ lines near $1.3 \, \mu$m, and
     the dark sky continuum (including solar-spectrum Zodiacal light) are
     summed to the thermal spectrum to account for those
     components. This model reproduces measurements like those of
     \citet{Sullivan2012} averaged over broad bands.  The \lamp\
     bandpass is situated in a narrow minimum in the emission spectrum
     (red line; see Section \ref{ssS:optics}), reducing the effect of
     variation in the OH lines on the overall
     brightness.}  \label{fig:lampatmosphere}
\end{figure*}

In this work, we investigate the stability of the large scale spatial
structure measured in a $\lambda / \Delta \lambda = 320$ band centered
at \lamplambda.  This wavelength is selected to lie between emission
lines in the atmosphere, as shown in Figure \ref{fig:lampatmosphere}.
We use a custom-built instrument, the Lyman Alpha Mapping Prototype
(\lamp), to measure the atmospheric stability in the \lamplambda\
window in a $1^{\circ}.2 \times 1^{\circ}.2$ field from Table Mountain
Observatory, California.  To characterize the stability of the
atmosphere, we investigate the total sky brightness, the noise in the
instrument, and the spatial power spectrum of the noise.  The \lamp\
instrument is presented in Section \ref{S:instrument}, the
observations and data analysis are presented in Sections
\ref{S:observations} and \ref{S:dataanalysis}, respectively, and
various results are shown in Section \ref{S:results}.  We conclude
with a discussion of the implications of these measurements in Section
\ref{S:discussion}.

\section{The Instrument} 
\label{S:instrument}

\lamp\ comprises of a modest commercial telescope coupled to a cryogenic
camera that images on a HAWAII-1 infrared detector array.  This
section describes (i) the \lamp\ instrument, including its optics,
mechanical assembly and readout electronics, and (ii) the
instrument's expected sensitivity to fluctuations.  In addition to
customized components, \lamp\ makes use of existing flight spares and
laboratory testing apparatus for the imaging camera of CIBER (Cosmic
Infrared Background Experiment; \citealt{Bock2013}, \citealt{Zemcov2013}).

\subsection{Instrument Description}

\subsubsection{The Optics}
\label{ssS:optics}

\begin{figure*}[p]
  \centering
  \resizebox{0.55\textheight}{!}{\includegraphics{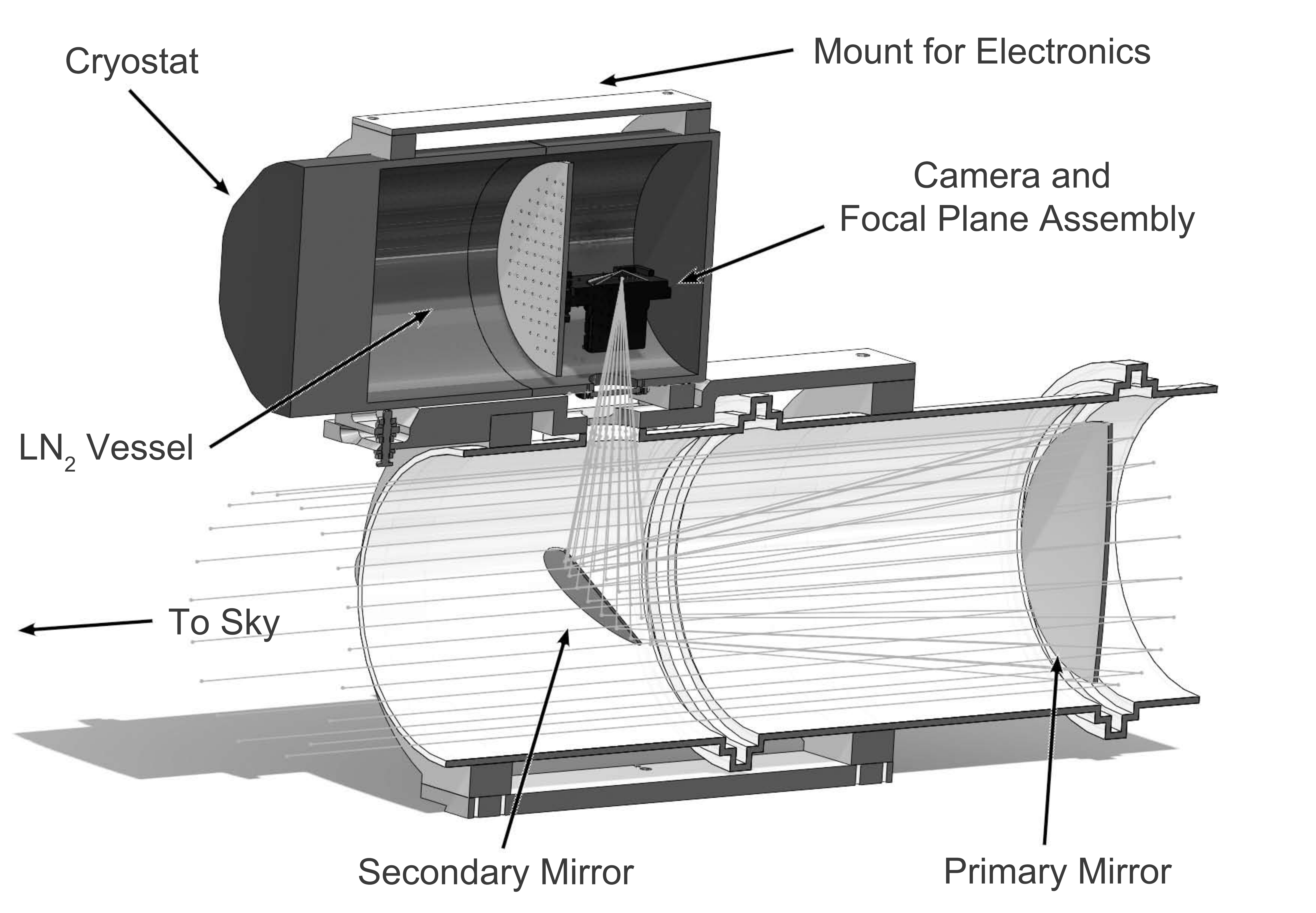}}
  \resizebox{0.55\textheight}{!}{\includegraphics{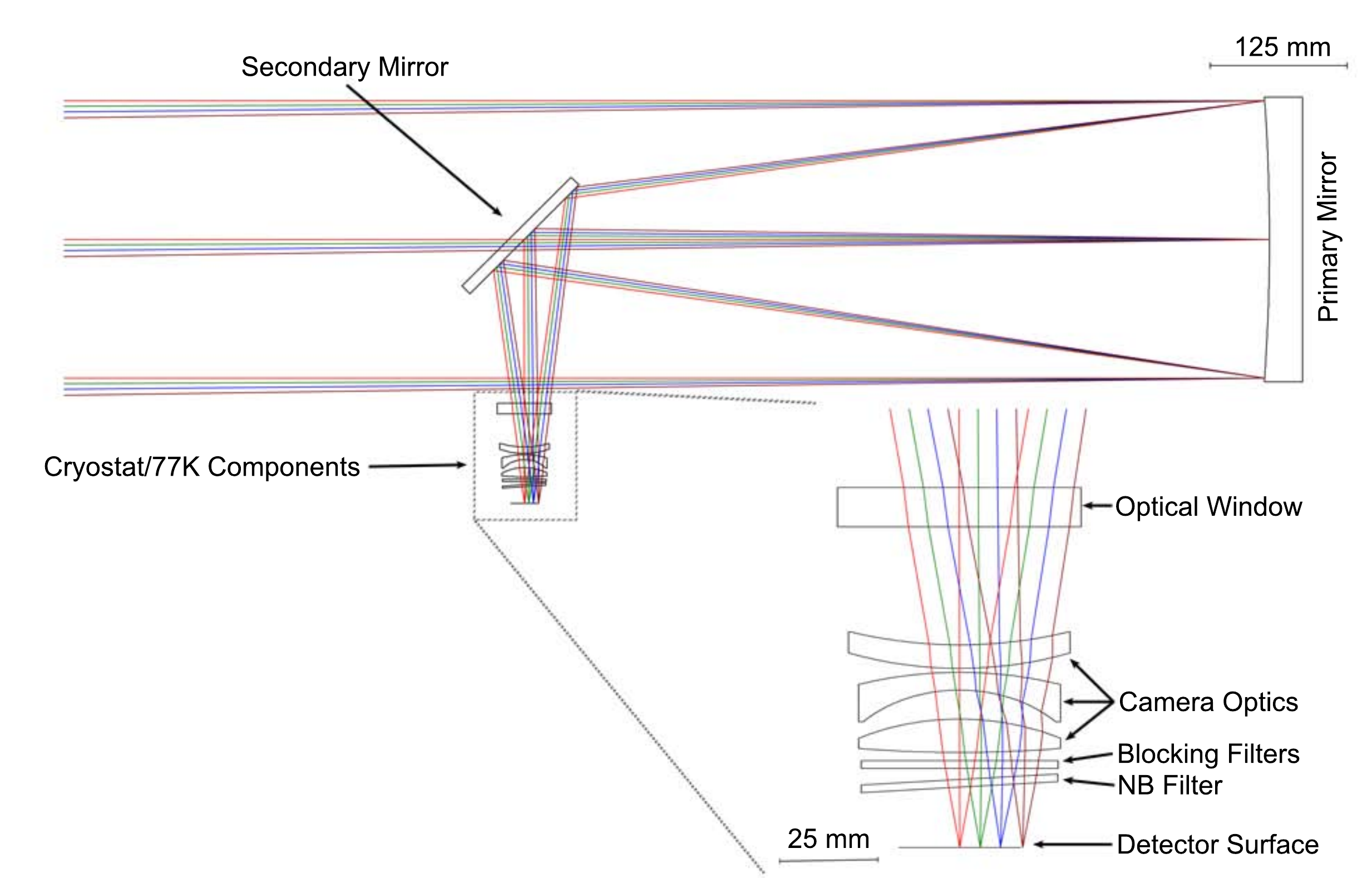}}
   \caption{Schematic views of the \lamp\ instrument and optical
     chain.  The upper panel shows a sectional view of a solid model
     of the \lamp\ system, showing the Newtonian telescope and
     cryostat including the camera optics and HAWAII-1 detector array.
     A narrow-band interference filter installed between the camera
     and detector is used to observe between the \oh\ airglow lines
     emitted by the Earth’s upper atmosphere.  The entire assembly is
     mounted on the top of TMF $24^{\prime \prime}$ telescope (see
     Figure \ref{fig:lampphoto}).  The lower
     panel shows a schematic of the LAMP optical chain highlighting
     the Newtonian telescope comprising a spherical primary and flat
     secondary, coupled to a wide field camera.  The optical ray trace
     shows the principal ray (red) and rays every $d \theta =
     +0.33^{\circ}$ (green, blue, brown). The negative $d \theta$ rays
     are suppressed but are symmetric to the positive rays.  The
     camera comprises 3 powered lenses and facility for up to 3
     optical filters in series.  The filters are tipped with respect
     to one another to eliminate reflections from the detector surface
     which would cause optical ghosting.  The optical efficiencies of
     the various components are given in Table \ref{tab:opticaleff}. 
     \label{fig:lampschematic}}
   \end{figure*}

\lamp\ uses a commercial $10^{\prime \prime}$ Newtonian
telescope\footnote{Manufactured by Parks Optical Inc.,
  \url{http://www.parksoptical.com}.} coupled to a liquid Nitrogen
(LN$_{2}$) cooled camera operating at $80 \,$K.  Figure \ref{fig:lampschematic} 
contains schematics of both the telescope and optical chain.  Light enters the camera
cryostat\footnote{Manufactured by IR Labs Part Number HDL-8
  \url{http://http://www.infraredlaboratories.com}.} through a $d=60
\,$mm optical window\footnote{Manufactured by Omega Optical Inc.,
  \url{http://www.omgeafilters.com}.} and is imaged by three aspheric
lenses\footnote{Manufactured by The Genesia Corporation,
  \url{http://www.genesia.co.jp}.} optimized to mitigate coma from the
telescope in order to achieve the full $1^{\circ}.2 \times 1^{\circ}.2$ field
of view (FOV).  The light is then filtered by a configurable
optical filter stack, and finally is detected by the near-IR detector array.
The window and lenses are all anti-reflection (AR) coated; the
transmissivity of the various optical components at $1190 \,$nm
determined either during manufacture or under test in the laboratory
are given in Table \ref{tab:opticaleff}.

\begin{table}[h]
\centering
\caption{\lamp\ optical efficiency budget at 1190 nm. \label{tab:opticaleff}}
\begin{tabular}{lc}
\hline
Component & $\eta$ \\ \hline
Mirrors & 0.90 \\
Window & 0.95 \\
Optics & 0.89 \\ \hline
Optics Total & 0.76 \\ \hline
Science \lamplambda\ Filter & 0.75 \\
Blocking Filter 1 & 0.76 \\
Blocking Filter 2 & 0.72 \\ \hline
Filter Total & 0.41 \\ \hline \hline
Total Optical Efficiency & 0.31 \\ \hline
\end{tabular}
\end{table}

\begin{figure*}
\centering
\resizebox{\hsize}{!}{\includegraphics{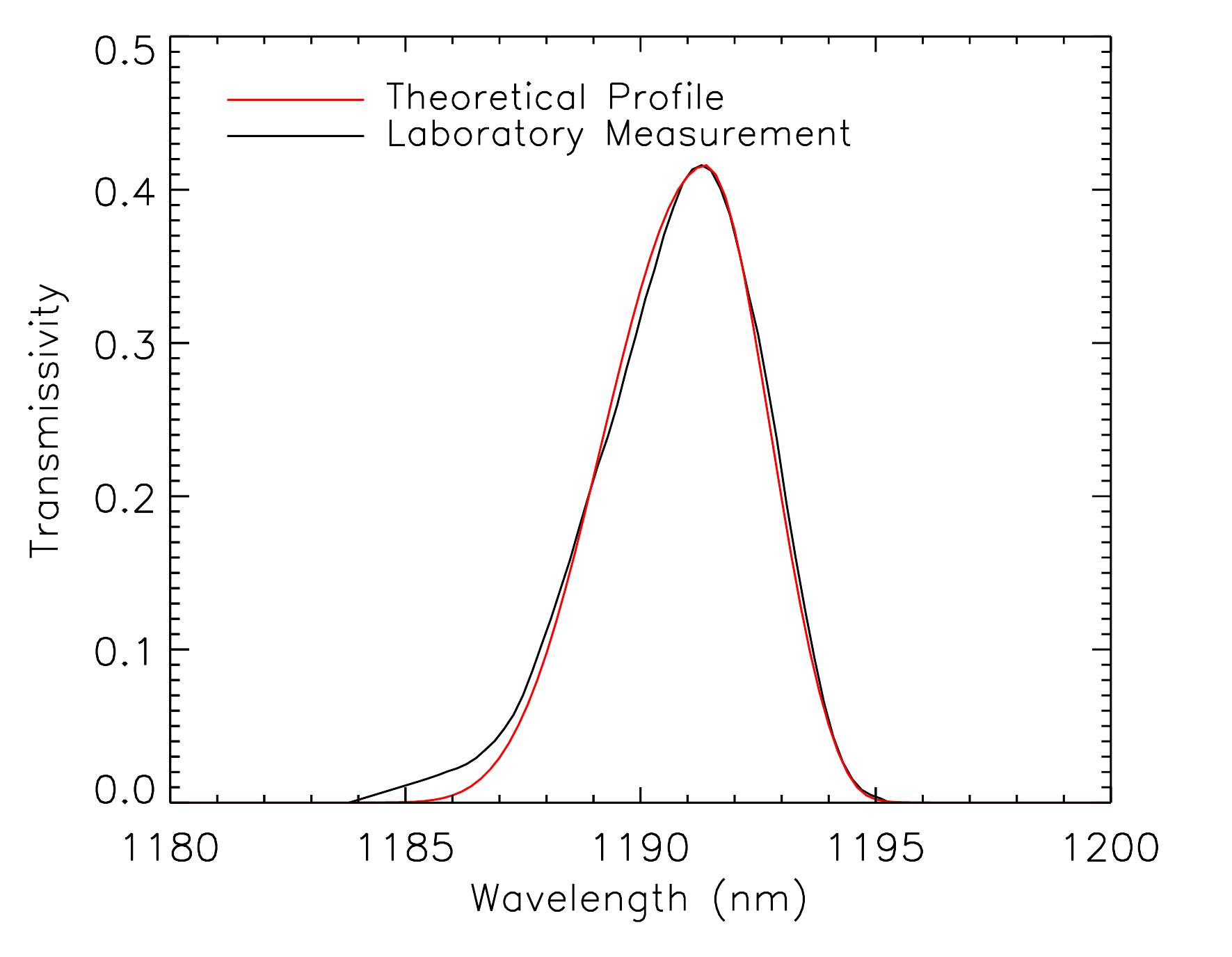}}
\caption{The transmissivity of the total \lamp\ filter stack,
  comprising a bandpass filter and two tilted out-of-band blocking
  filters.  The measured transmissivity is determined in the
  laboratory at $77 \,$K using a monochromator with a $\delta \lambda
  = 1.8 \,$nm dispersion, causing a broadening of the measurement
  compared to the intrinsic width of the filter.  The theoretical
  transmissivity curve is calculated from the $\theta = 0^{\circ}$
  angle of incidence filter transmissivity provided by the filter
  manufacturer in two steps.  First, the transmissivity curve is
  convolved with the square spectral function of the monochromator.
  Secondly, the transmissivity short-ward of the filter peak is
  convolved with a ``blue broadening'' function following $\lambda =
  \lambda_{0} \cos \theta$ \citep{Korngut2013}, where $\lambda_{0}$ is
  the response at normal angle of incidence and the optics are $f/3.4$
  through the bandpass filter.  The FWHM of this filter configuration
  as measured is 4.1 nm, but after removing the effect of the
  monochromator spectral function the intrinsic width of the filter
  stack is determined to be $3.7 \,$nm.
  \label{fig:lampfilters}}
\end{figure*}

The filter stack is designed to accomodate up to 3 optical filters in
series.  In the configuration used for this measurement, we installed
both the narrow band \lamplambda\ filter and two blocking filters to
reduce the out of band transmissivity of the
system\footnote{Manufactured by Chroma Technology Corp.~to customized
  specifications, \url{http://www.chroma.com}}.  This stringent
blocking of the out of band photons is crucial, motivated by the fact
that the airglow has a specific surface brightness of $I_{\lambda}
\sim 5000 \,$\nw\ $\mu$m$^{-1}$ in the near IR, so that filter leaks
could easily dominate the small in-band signal in this measurement.
The filters are installed with the narrow band filter closest to the
camera optics parallel to the detector surface, and the two blocking
filters closest to the detector are tipped by $4^{\circ}$ to
mitigate optical ghosting from reflections from the detector surface.
In Figure \ref{fig:lampfilters} we show the theoretical spectral
bandpass for the instrument, and measurements of the bandpass made in
the laboratory.  We place upper limits on the out of band blocking
using laboratory measurements, and find that for $0.8 < \lambda < 2.5
\, \mu$m the out of band rejection is $> 10^{4} \,$.
This gives an upper limit on the out of band contribution to the
photocurrent of $< 1.5 \,$\%, integrating over the wavelength range to
which our HgCdTe detectors are responsive.

\subsubsection{Mechanical and Cryogenic Systems}

The telescope tube is firmly supported by a pair of circular clamps made
of cast-iron.  The top and bottom of these clamps are mounted on metal
interfaces.  The top interface is used to attach a small cryostat
that houses the camera lens and its detector array.  The cryostat
holds approximately 5 liters of LN$_{2}$, which lasts for 48 hours
between service.  The cryogen passively cools both the camera lenses
and detectors to $<$ 80 K.  The entire instrument assembly, totaling
a weight of 200 lbs, is mounted on an existing 60 cm telescope (at Table Mountain Observatory), to benefit from the use of the larger telescope's pointing and tracking.

The \lamp\ system uses a focal plane assembly (FPA) from the CIBER
instrument, comprising nested light-tight housings for the detectors
and associated cryogenic electronics allowing active thermal control
of the detector.  The FPA unit is described in detail in
\citet{Zemcov2013}.  The FPA unit is attached to the camera at a
distance determined to bring the telescope in focus in the laboratory.

\subsubsection{Detector and Readout Electronics}

\lamp\ uses the same $1024 \times 1024$ HAWAII-1\footnote{Manufactured
  by Teledyne Scientific \& Imaging LLC,
  \url{http://www.teledyne-si.com}.} HgCdTe detector array as is used
by the CIBER instrument\footnote{Serial number Hawaii-211.}.  A
detailed presentation of the properties of this detector can be found
in \citet{Bock2013}.  \lamp\ also uses the same readout chain as used
for CIBER, slightly modified to operate with a single channel.  The
electronic system is presented in Section 4 of \citet{Zemcov2013}; for
\lamp\, the system is used in `Serial Mode'.  Because of their
heritage of use in sounding rockets where there are concerns about
targeting computer trigger loss, the electronics are configured to
have a maximum of 70 integrations before resetting the array.  This
limits the maximum data set length obtainable with these electronics
to $< 110 \,$s.

\subsection{Theoretical Sensitivity}
\label{sS:theorysens}

We can calculate the theoretical sensitivity of \lamp\ given known
instrument parameters and the approximate brightness of the sky.  The
equation relating the photo current at the detector $i_{\rm{phot}}$ to
the brightness of a beam-filling astronomical source $\lambda
I_{\lambda}$ is:
\begin{equation}
\label{eq:telescope}
i_{\rm{phot}} \simeq \lambda I_{\lambda} \left( \frac{\eta A \Omega}{h
    \nu}\frac{\Delta \lambda}{\lambda} \right),
\end{equation}
where $\eta$ is the system efficiency, $A$ is the area of the
aperture, $\Omega$ is the angular size of a pixel, $\Delta \lambda / \lambda$
is the fractional filter width, and $h \nu$ is the energy of the
photons in the \lamp\ band (see the Appendix to \citealt{Bock2013} for
a discussion).  The photometric surface brightness calibration
$\mathcal{C} = \lambda I_{\lambda} / i_{\rm{phot}}$ is given by the
term in brackets on the right hand side of Equation
\ref{eq:telescope}, and relates the sky brightness in \nw\ to the
measured photocurrent in \eps.  The instrument parameters we have
determined for LAMP are summarized in Table \ref{tab:instparams}.
Based on these values, we calculate $\mathcal{C} = 5.65 \times 10^{4}
\, \text{nW m}^{-2} \text{ sr}^{-1} / e^{-} \text{ s}^{-1}$.

\begin{table}[ht]
\centering
\caption{LAMP Instrument Parameters \label{tab:instparams}}
\begin{tabular}{lcc}
\hline
Parameter & Value & Units \\ \hline
Operating Wavelength & \lamplambda & nm \\
Filter Width & 3.7 & nm \\
F\# & {3.4} & \\
Focal Length & 865.9 & mm \\
Clear Aperture & $390$ & cm$^{2}$ \\
Pixel Size & $4.3 \times 4.3$ & arcsec \\
Field of View & $1.2 \times 1.2$ & deg \\
Optical Efficiency & 0.31 & \\
Array QE & 0.52 & $\ast$ \\
Total Efficiency & 0.16 & \\
Array Format & $1024^{2}$ & \\
Pixel Pitch & 18 & $\mu$m \\
Read Noise (CDS) & 10 & $e^{-}$ \\
Frame Interval & 1.78 & s \\ 
Theoretical Surface Brightness Calibration & $5.65 \times 10^{4}$ & \nw/\eps \\
\hline
\multicolumn{3}{l}{$\ast$ Array QE is estimated from QE measured at $2.2 \,
  \mu$m and scaled based on} \\
\multicolumn{3}{l}{the response of a typical Hawaii-1.} \\
\end{tabular}
\end{table}

In this analysis, we fit a linear model to constant frame interval
reads of the detector array to estimate the photocurrent in a given
integration.  The instrument read noise $\sigma_{\rm{read}}$ for this
estimator is given by:
\begin{equation}
\label{eq:readnoise}
\sigma_{\rm{read}}^{2} = \frac{12 N \sigma_{\rm{CDS}}^{2}}{(N^{2}-1) T_{\rm{int}}^{2}} 
\end{equation}
where $N$ is the number of frames in an integration,
$\sigma_{\rm{CDS}}$ is the single frame read noise estimated using the
root mean squared variation in a correlated-double-sample, and
$T_{\rm{int}}$ is the integration time \citep{Garnett1993}.  In this
analysis, full integrations have a $1 \sigma$ read noise of
$\sigma_{\rm{read}} = 43 \,$m\eps, corresponding to $2430 \,$\nw.

The photon noise in this measurement is given by:
\begin{equation}
\label{eq:photonnoise}
\sigma_{\rm{photon}}^{2} = \frac{6 F (N^{2}+1)}{5 T_{\rm{int}}
  (N^{2}-1)}
\end{equation}
where $F$ is the measured surface brightness at the detector.  In
these data, the typical surface brightness $F \approx 95 \,$m\eps,
yielding a typical $\sigma_{\rm{photon}} = 33 \,$m\eps, which
corresponds to $1840 \,$\nw.  The data presented here are read noise
dominated by a factor of $1.3$.  Equations \ref{eq:readnoise} and
\ref{eq:photonnoise} predict that the measurement would have equal
contributions from read and shot noise after $\sim 140 \,$s of
integration time.

In this work we compute angular power spectra $C_{\ell}$ of images to
investigate the spatial stability of the noise with time.  Assuming
uncorrelated white noise and ignoring sample variance, the power
spectrum uncertainty $\delta C_{\ell}$ is given by:
\begin{equation}
\delta C_{\ell}  = \sqrt{\frac{2}{f_{\rm sky} (2 \ell + 1) {\frac{\Delta
    \ell}{\ell} }}} (\sigma_{\rm pix}^{2} \Omega_{\rm pix} e^{\theta_{\rm beam}^{2} \ell^{2}}),
\end{equation}
where $\Delta \ell$ is the band power bin width, $\sigma_{\rm pix}$ and
$\Omega_{\rm pix}$ are the surface brightness rms noise and solid angle
of each pixel, respectively, and $\theta_{\rm beam}$ is the point
spread function width \citep{Knox1995}.

\section{Observations}
\label{S:observations}

\begin{figure*}
\centering
\resizebox{\hsize}{!}{\includegraphics{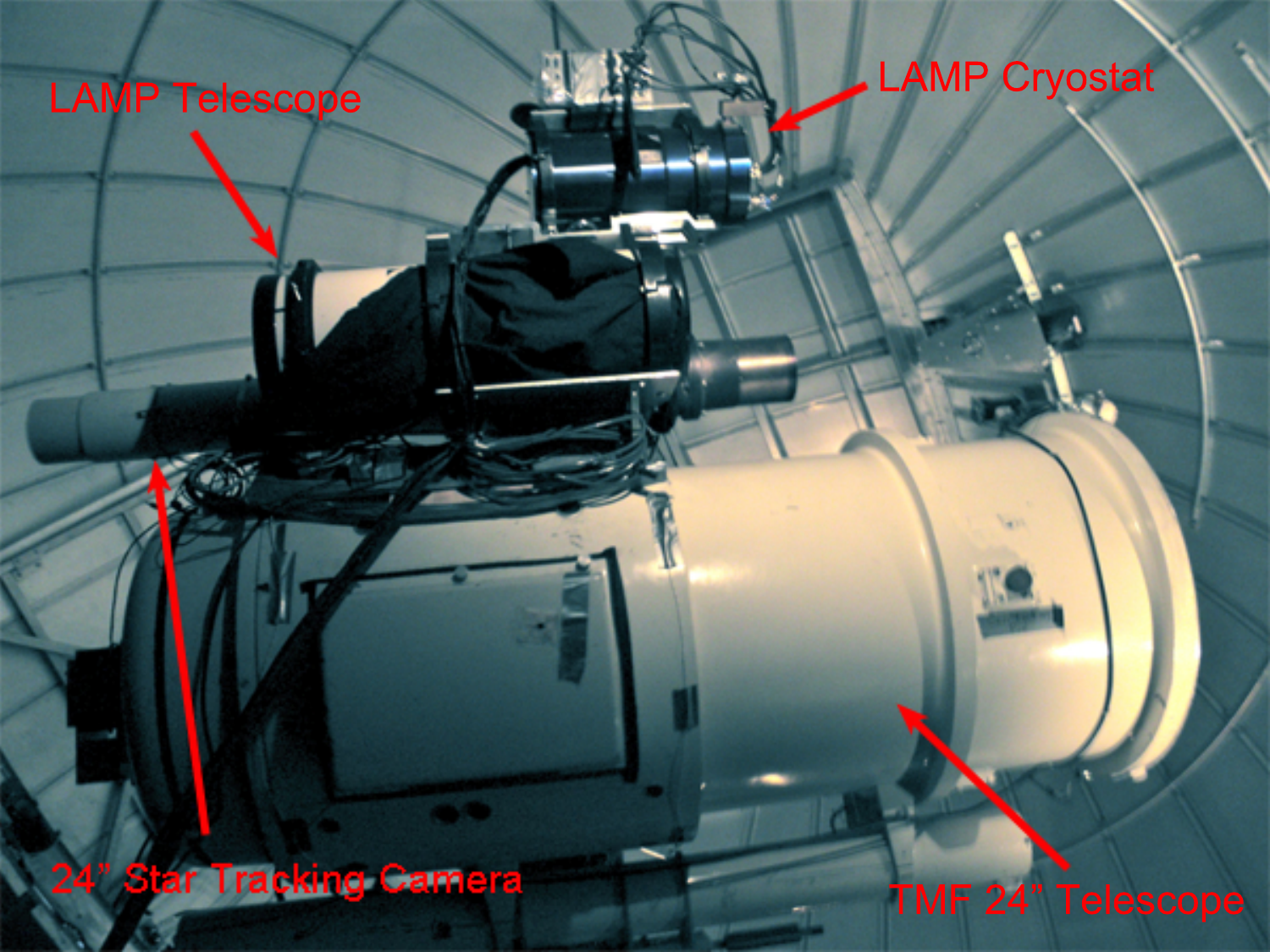}}
\caption{A photograph of \lamp\ mounted on the $24^{\prime \prime}$
  TMF telescope.  The \lamp\ assembly is mounted to the larger
  telescope using an existing interface bracket.  After installation,
  we verified the $24^{\prime \prime}$ telescopes's pointing and
  tracking system, which to the level we were able to measure
  performed nominally under the additional load of the \lamp\
  assembly. \label{fig:lampphoto}}
\end{figure*}

We conducted our astronomical observations from the Jet Propulsion
Laboratory's Table Mountain Test Facility (TMF) Astronomical
Observatory\footnote{Longitude 117.7W, Latitude 34.4N, Altitude $2285
  \,$m asl.}.  As shown in Figure \ref{fig:lampphoto}, \lamp\ was
mounted on top of the $24^{\prime \prime}$ telescope\footnote{The TMF
  $24^{\prime \prime}$ telescope is an Astro-Mechanics
  Ritchey-Chretien reflector on an off-axis German equatorial mount,
  see \url{http://tmoa.jpl.nasa.gov} for more information.}.  Doing so
allows \lamp\ to use the existing pointing and tracking mechanism of
the $24^{\prime \prime}$ system.  Because \lamp's optical axis was
offset from the $24^{\prime \prime}$ telescope's bore sight a small
tracking error was introduced, an effect which we quantify in Section
\ref{sS:psf}.

\begin{figure*}[p]
\centering
\resizebox{\hsize}{!}{\includegraphics{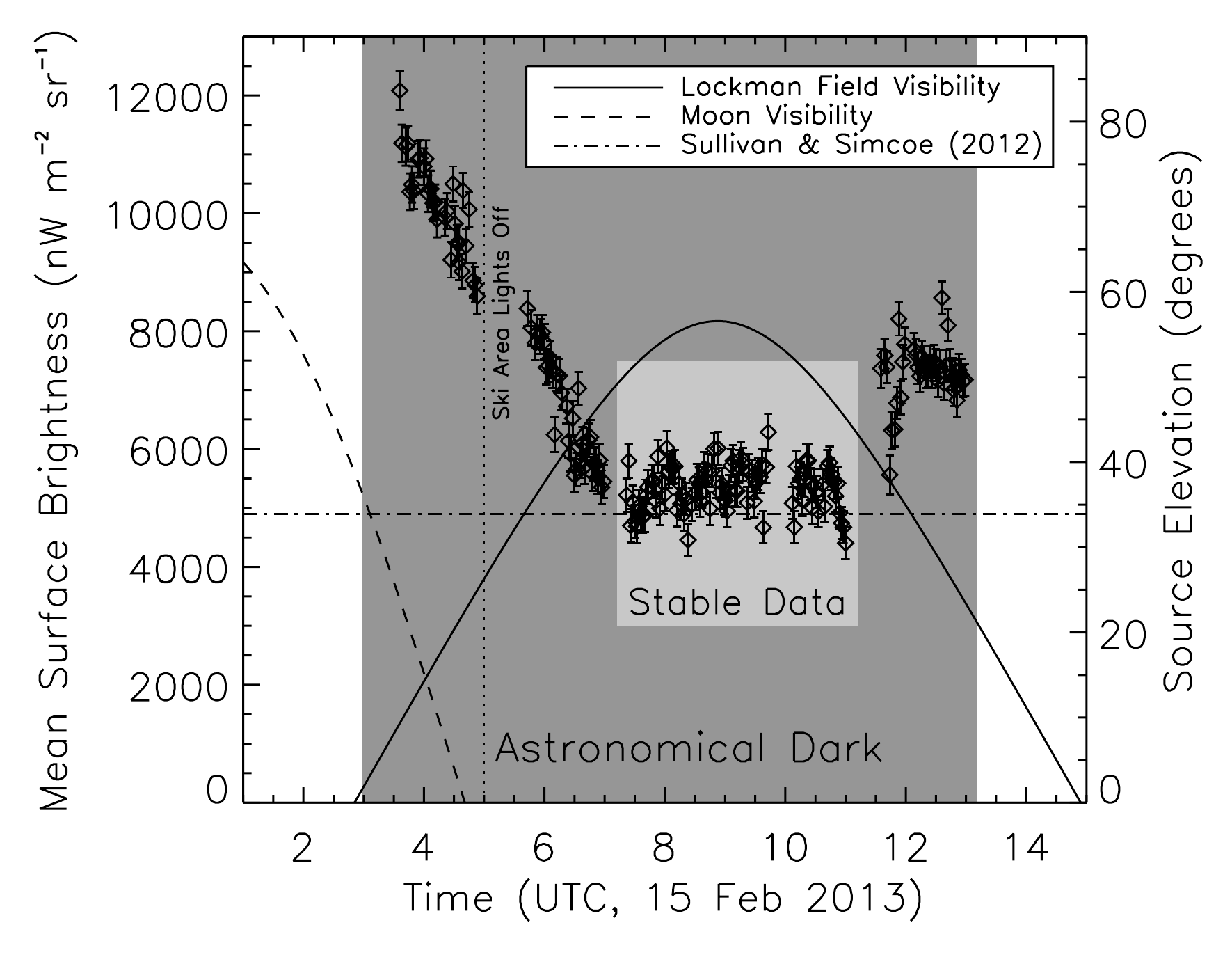}}
\caption{The derived near IR sky brightness as measured by \lamp\
     on the night of 15 Feb, 2013 UTC from the Table Mountain
     Observatory.  The astronomically dark period is indicated by the
     shaded area.  The Moon (dashed line) set at approximately 04:30
     UTC as the target field was rising (solid line).  The TMF
     Observatory is located close to a skiing area whose Fluorescent
     lights are turned off at 05:00 UTC (22:00 local time).  Airglow
     emission which is a function of the Lockman field's elevation is
     observed before and after a stable plateau in the sky surface
     brightness between 07:30 and 11:00 UTC (labeled ``stable data'').
     These data are used in the stability study we present in this
     report.  The small gap in the stable data near 09:30 are due to
     electronics errors and results from these integrations are not
     reported.  The mean surface brightness of the sky is $5330 \,$\nw,
     corresponding to mag$_{\rm{AB}}= 19.7$, approximately $480 \,$\nw
     of which can be attributed to Zodiacal light \citep{Kelsall1998}.
     This is close to the sky brightness at \lamplambda\ reported by
     \citet{Sullivan2012} (dot-dashed line), measured by the
     FIRE instrument at an altitude of $2380 \,$m from Las Campanas
     Observatory, Chile.
     \label{fig:skybrightness}}
\end{figure*}

Sky observations were recorded for 03:30 to 13:00 during the night of
February 15th, 2013 (UTC).  Local astronomical darkness\footnote{See
  \url{http://aa.usno.navy.mil/data/docs/RS\_OneYear.php}.} began
around 03:00 UTC, and a waxing crescent moon of $\approx \, 33 \,$\%
illumination was present at low elevation early in evening, setting
around 04:30 UTC.  After initial pointing calibrations, we observed the
SWIRE-Lockman\footnote{\url{http://irsa.ipac.caltech.edu/data/SPITZER/SWIRE/}.}
Hole field centered at right ascension $10^{\rm{h}}45^{\rm{m}}$,
declination $58^{\circ}00^{\prime}$ (referred to hereafter as the
``Lockman'' field).  The data on which we report were recorded with
the Lockman field at 45 degrees or more above the horizon ($< 1.4$
airmasses) over the night.  The seeing was $\sim 2.5^{\prime \prime}$
throughout the night.  Figure \ref{fig:skybrightness} shows the
visibility of the Lockman field, the moon, and various other events
throughout the observation period.

A total of 245 on-sky integrations were recored, with 232 of these
having $T_{\rm{int}} > 105 \,$s.  Of these, a large number were
recorded during the early evening and early morning when the target
field is low in the sky, and there are dynamic sources of emission
beyond our control before midnight local time (for example, the
fluorescent lighting and cars headlights associated with a local ski
area).

The mean brightness of the fields reaches a constant plateau of $5330
\pm 30 \,$\nw\ between 07:15 and 11:15 UTC over which no discernable
time gradient is present.  We call this set of 107 integrations
``stable period'' data.  Of these, 10 integrations must be excluded as they
do not meet the minimum $T_{\rm{int}}$ requirement, 5 must be
excluded due to electrical pickup problems, and one must be excluded
due to a tracking error.  This leaves a set of 91 integrations which
are sufficiently stable for analysis.

Dark frames are measured with the opening of LAMP telescope covered
using a customized light-tight aperture cover.  Data are acquired at
regular intervals over the evening to monitor changes in the dark
current.  We measure an elevated dark current at the very beginning of
the evening while the sky is darkening, likely due to light leaks and
instabilities in the system.  Over the course of the evening, the dark
current drops to $\sim 0.25 \,$\eps\, and is stable before and after
the stable period.  Approximately twenty dark integrations bracket the
stable data period, and we use the mean and ensemble variance of these
to be the dark current correction and error in that estimate.

Flat fields were measured during twilight, at the beginning and the end of
the night.  We take
care to use only flat field frames in which the sky gives $S/N > 100$
per pixel but is not bright enough that non-linearities in the
detector would become problematic, giving approximately twenty flat
fields from which to compute the flat field correction.

\section{Data Analysis}
\label{S:dataanalysis}

\subsection{Low Level Data Analysis}

The low level data reduction follows that presented in
\cite{Zemcov2014}.  For each integration, lines are fit to array reads
3 to 60, yielding a constant integration time of $58 \, \rm{frames}
\times 1.78 \,$s/frame = $104 \,$s.  The first two array reads are not
used in the fit to reduce susceptibility to electrical transients
associated with array resets.  The resulting photocurrents are
calibrated to \eps\ using known calibration factors which depend on
the detector and electronics.  Raw photocurrents of 0.25 \eps\ are
typical for these data.

After the photocurrent estimation, we correct for the dark current.
We use the dark data discussed in Section \ref{S:observations} in each
frame, where the photocurrent is estimated in the same way as for the
sky integrations.  The best dark current estimate is formed by
computing the weighted mean of the individual dark frames that bracket
the stable period integrations.  We chose a weight function that
weights according to the time difference between the sky integration
being corrected and each dark frame $t_{i-j}$ according to:
\begin{equation}
d_{i} = \sum_{j} t_{i-j}^{-1} d_{j} / \sum_{j} t_{i-j}^{-1}
\end{equation}
where $d_{i}$ is the dark current estimator for sky integration $i$,
and $d_{j}$ are the $j$ individual measurements of the dark current.
The effect of misestimating the dark current correction on our results
is investigated in Section \ref{sS:systematics}.

We generate a uniform-illumination responsivity response correction
(``flat field'') using the dedicated measurements presented in Section
\ref{S:observations}.  To generate the flat field, we compute the sum
of all eleven flat field measurements with photocurrents $100 < I_{q}
< 350 \,$\eps\ to ensure detector linearity, and then divide by the
sum of the median of each of the measurements.  This weighting scheme
is optimal in the limit that the integrations are dominated by photon
noise, as is the case for $I_{q} > 50 \,$\eps, and yields a
measurement of the flat field which is normalized to have unity
median.  The effect of a systematic error in the estimation of the
flat field correction on on our results is investigated in Section
\ref{sS:systematics}.

Next, the images must be registered to absolute astrometry to allow
masking of known sources.  This calculation is performed using the
{\sc astrometry.net}\footnote{\url{http://astrometry.net}} code which
automates astrometric registration on arbitrary astronomical
instruments \citep{Lang2010}.  The accuracy of the astrometric
solution is evaluated by computing the difference between the centroid
of a subset of bright catalog sources and their positions given by the
astrometric solution.  This calculation is performed for all suitably
bright sources in the image, producing sets of differences between the
centroids and astrometry solution positions in both axes. For a given
image we define the astrometric pointing error to be the standard
deviation of each set of differences for each integration. To produce
global values over the observing period, the above process was
repeated for all stable period integrations.  To determine the
pointing error over the stable period, we fit a linear model to the
time-ordered set of standard deviations in each axis. As a
conservative estimate, the maximum values of the line fits over the
stable period are taken to be the overall pointing error.

We calculate the overall pointing error as the mean ellipticity of an
ellipse where the mean offset in right ascension and declination
compose the semi-major and semi-minor axes.  The total uncertainty is
$0.9^{\prime \prime}$ in right ascension and $1.2^{\prime \prime}$
in declination, yielding a total astrometric uncertainty of
$1.0^{\prime \prime}$.

The \lamp\ photometric calibration is calculated using stars from the
2MASS catalog \citep{Skrutskie2006}.  Aperture photometry is performed
on the twelve stars with $6 < J < 9.5$ in the \lamp\ images using a
summing aperture of 10 pixels.  The best fitting linear relation
between the \lamp\ photometry and the known flux of the sources is
then determined, yielding a calibration factor $\mathcal{C} = 5.52
\times 10^{4} \,$\nw / \eps.  We limit the uncertainty in the
calibration to be $2.5 \,$\% using the nominal fitting error in the
scaling between the \lamp\ photometry and 2MASS fluxes.  The 2MASS
photometric uncertainty is $\sim 1 \,$\%, giving a total photometric
calibration uncertainty of $\sim 3 \,$\%.  The calibration factor
agrees with the theoretical calculation (listed in Table
\ref{tab:instparams}) within uncertainties.

\begin{figure*}[p]
   \centering
   \resizebox{\hsize}{!}{\includegraphics{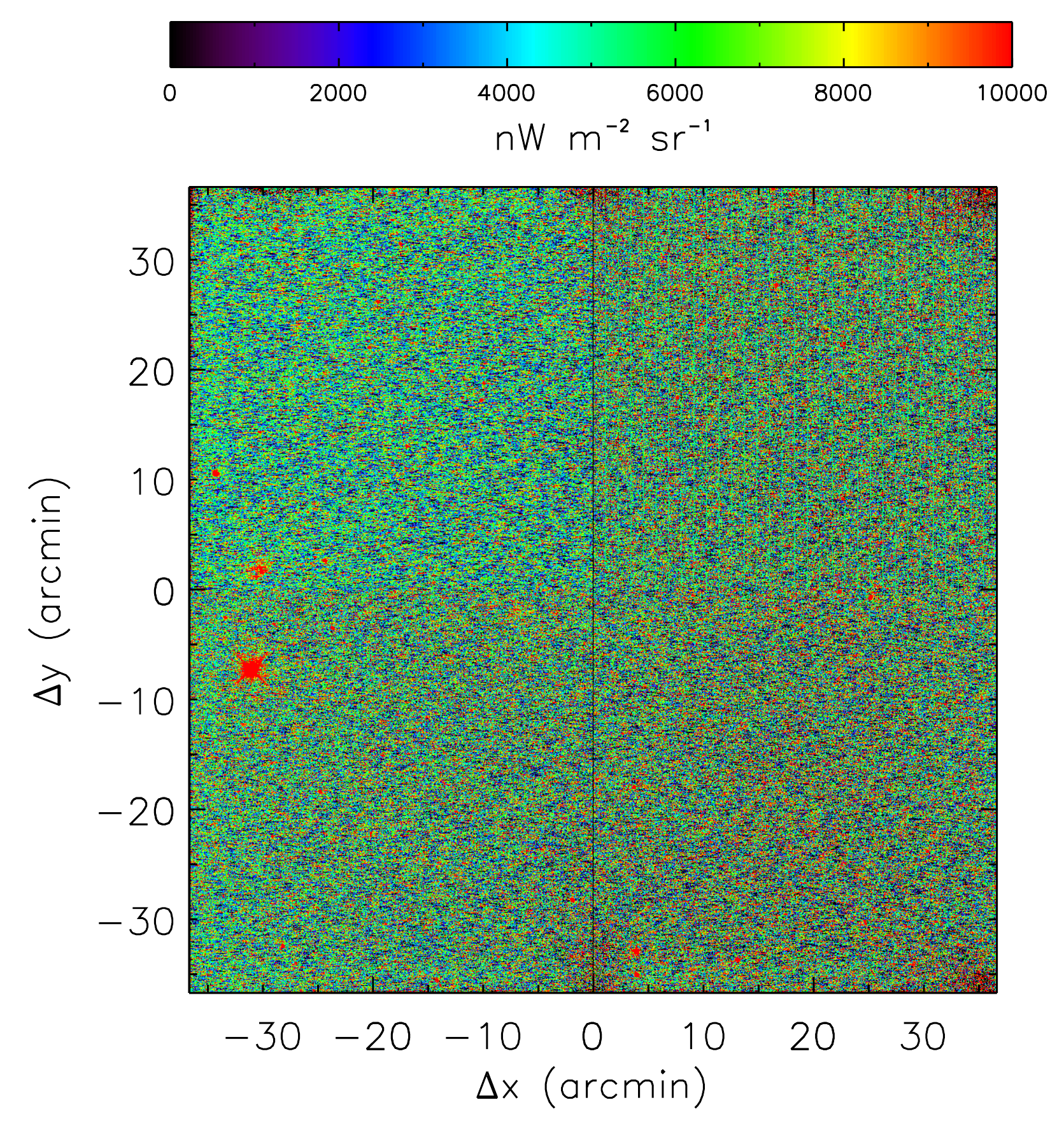}}
   \caption{A raw $T_{\rm{int}} = 104 \,$s integration on the Lockman
     field at \lamplambda\ as seen by \lamp.  The image has had dark
     current subtraction and flat field responsivity correction
     applied.  The pixel size is $4.3^{\prime\prime}$ on a side,
     giving a field of view of $1.5$ square degrees centered at
     $10^{\rm{h}}47^{\rm{m}}17^{\rm{s}}$,
     $57^{\circ}53^{\prime}30^{\prime\prime}$.  Several features are
     apparent in this image, including an image of the $J=3.13$ star
     HD 93132 near $(-32,-7)$ with diffraction spikes from the
     secondary mirror support, a reflection of its image near
     $(-32,2)$, pixels with large multiplexer glow and fabrication
     defects around the edges of the array, and unresponsive columns
     in the upper right hand quadrant.  These features, as well as
     bright stars and galaxies in the 2MASS catalog, are masked later
     in the analysis.
     \label{fig:imageunmasked}}
   \end{figure*}

As an example, we show an image to which the full data processing has
been applied in Figure \ref{fig:imageunmasked}.  In order to calculate
power spectra from these images, the instrumental artefacts and images
of stars must be masked, which in turn requires knowledge of the point
spread function.

\subsection{Point Spread Function and Source Masking}
\label{sS:psf}

\begin{figure}[htp]
\centering
\resizebox{\hsize}{!}{\includegraphics{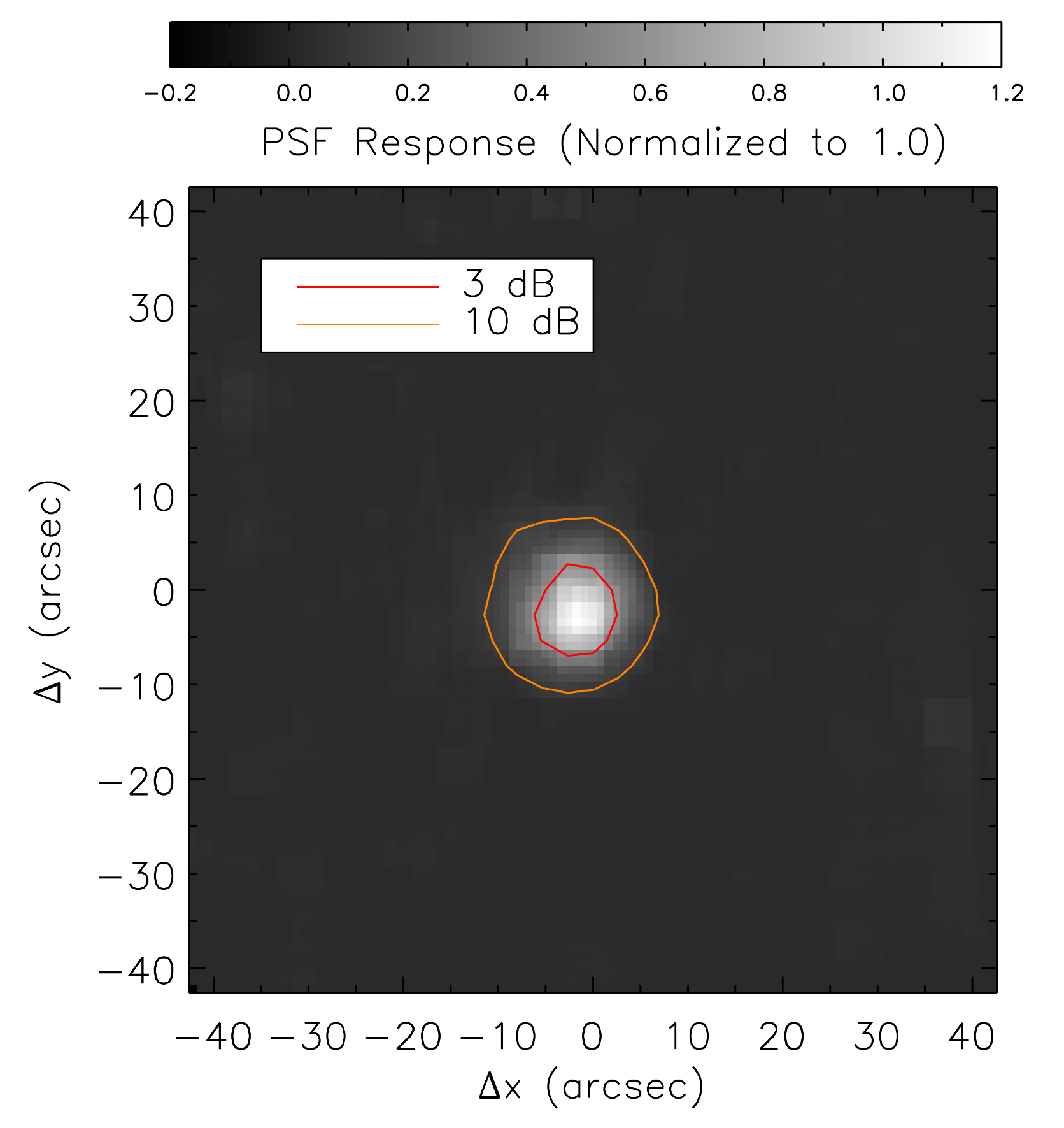}}
\caption{The \lamp\ PSF measured in a single integration on a
  $3\times$ sub-sampled pixel grid.  To measure the PSF, we combine
  stacks on all stars with $5 < J < 10$ for the extended PSF, and $10
  < J < 11$ for the core PSF, which yields a determination of the
  overall PSF sufficient to allow accurate source masking.  The
  contours show the $3 \,$dB (i.e.~the FWHM) and $10 \,$dB contours of
  the PSF smoothed to \lamp's native
  pixelization.  \label{fig:psfcontour}}
\end{figure}

The \lamp\ point spread function (PSF) is estimated using a stack of
sources similar to the method presented in \citet{Zemcov2014}.  The
core PSD is determined by stacking all 2MASS $J-$band catalog sources
with $10 < J < 11$, and the extended PSF is determined by stacking all
2MASS sources with $5 < J< 10$.  The full PSF is computed by splicing
the two measurements together using a linear scaling in regions where
the PSF is between $1 \,$\% and $10 \,$\% of its peak in the core
stack.  An example of the PSF stack result in a single integration is
shown in Figure \ref{fig:psfcontour}.

For each integration, we estimate the width of the PSF by fitting a
two-dimensional Gaussian to the PSF stack.  The mean one-dimensional
PSF width is estimated from the average of the semi-major and
semi-minor widths of the best fitting result.  The mean FWHM value over
the entire data set is $8.5^{\prime \prime}$, but is variable with
time from $8.0^{\prime \prime}$ at the beginning of the stable data
region to as poor as $9.5^{\prime \prime}$ at the end.  It is not
clear what causes this degradation, but other than the $\sim 20 \,$\%
broadening in both axes no trend in ellipticity is detected.

The total PSF width is the quadrature sum of several contributions,
including the intrinsic PSF, the focus of the telescope, pointing
errors, and atmospheric seeing.  The contribution from
each of these contributions is summarized in Table
\ref{tab:psferrbudget}.  

\begin{table}[ht]
\centering
\caption{Point Spread Function Width Budget \label{tab:psferrbudget}}
\begin{tabular}{lc}
\hline
Component & Contribution (FWHM) \\ \hline
Intrinsic PSF Width & $6^{\prime \prime}.1$ \\
Pointing Errors & $1^{\prime \prime}.5$ \\
Atmospheric Seeing & $2^{\prime \prime}.5$ \\ \hline
Total Calculated PSF Width & $6^{\prime \prime}.8$  \\ \hline
Total Measured PSF Width & $8^{\prime \prime}.5$ \\ \hline
Inferred Defocus & $5^{\prime \prime}.1$ \\  
\end{tabular}
\end{table}

The instrument's intrinsic PSF FWHM is estimated as equal to
the full width of the 68 percent encircled energy contour in a ray
trace simulation.  Though the encircled energy diagram changes
slightly over the array, this contour is always less than $25 \,
\mu$m, corresponding to $6^{\prime\prime}.1$ at the $18 \, \mu$m pixel
pitch of the array. 

Over the course of an integration the instrument pointing may not
perfectly track the sky, either coherently with time (``smear'') or as
a random noise (``jitter''), and degradation of the PSF width could
result.  To constrain pointing smear, we compute the difference
between the astrometry solutions in two time-halves of the same full
integration, which in the absence of smearing should be zero.  The
astrometry solution is independently determined for each
half-integration, and the difference between them is computed.  This
procedure yields a mean difference of $1^{\prime \prime}.0$ over the
stable data period, with no evidence for ellipticity.  To constrain
pointing jitter, we calculate the difference between the known catalog
positions of all $5 < J < 10$ sources and their positions from the per
integration astrometry solution.  For each integration, we then
compute the mean difference over all sources, yielding a monitor of
the overall accuracy of the astrometric solution over time.  The mean
difference between the known and solved positions of catalog sources
over the stable data period is $\delta \theta = 1^{\prime \prime}.1$,
and we detect no significant difference between right ascension and
declination or time variation.  Assuming the pointing jitter and
smearing are uncorrelated so they may be added in quadrature, we
estimate the total pointing error to contribute $1^{\prime \prime}.5$
to the PSF width.

The atmospheric seeing was recorded as $2^{\prime \prime}.5$ during
the observation period.

The measured PSF width differs from the sum of the various
contributions by
$\sqrt{8^{\prime \prime}.5^{2} - 6^{\prime \prime}.8^{2}} = 5^{\prime
  \prime}.1$.
This is likely due to poor focus of the instrument, which was
mechanically set during laboratory testing and could not be verified
after mounting at TMF.  These could be discrepant due to alignment
errors during instrument mounting.

Having determined the PSF, the images can be masked.  The masking
algorithm follows that presented in \citet{Zemcov2014}, with
$\alpha_{m}=-8^{\prime \prime}.5 \,$mag$^{-1}$ and $\beta_{m} =
141^{\prime \prime}.0$, resulting in a slightly broader mask than used
for the CIBER data.  As in the \citet{Zemcov2014} analysis, these
parameters are determined by computing the power spectra of
simulations of the 2MASS sources in the image and estimating the
threshold at which the power from residual sources is significantly
less than the noise in the image.  In the \lamp\ analysis, we mask to
$J = 16 \,$mag, as the per integration surface brightness sensitivity
is significantly larger than this source flux.  Between astronomical
sources and masks of static structure on the detector typically $16
\,$\% of pixels are masked.

\begin{figure*}[p]
   \centering
   \resizebox{\hsize}{!}{\includegraphics{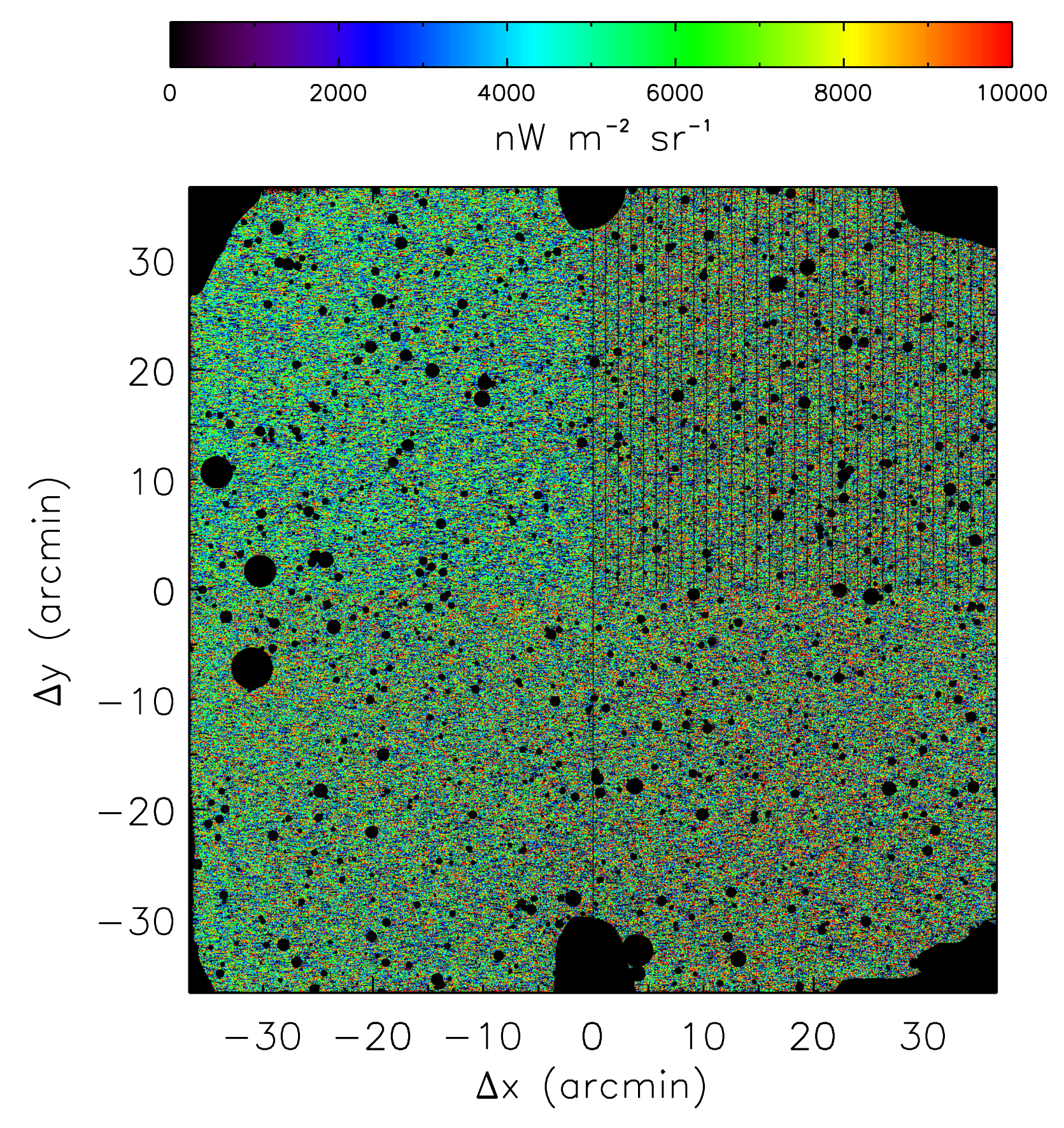}}
\caption{The same array integration as shown in Figure
  \ref{fig:imageunmasked}, but with masking of array defects and
  $J > 17.5$ astronomical sources applied.  Approximately $16 \,$\% of
  pixels are lost to the mask.  The resulting image
  structure is consistent with a combination of read noise and shot
  noise from the diffuse airglow emission.
  \label{fig:imagemasked}}
   \end{figure*}

\subsection{Power Spectrum Estimation}
\label{sS:powerspectra}

The spatial auto-power spectra are computed using a version
of the {\sc master} formalism \citep{Hivon2002} in which the true sky
$\widetilde{C_{\ell}}$ is related to the measured sky power $\langle
C_{\ell} \rangle$ by:
\begin{equation}
\label{eq:ps}
\widetilde{C_{\ell}} = \sum_{\ell^{\prime}} M_{\ell
    \ell^{\prime}}^{-1} \frac{(\langle  C_{\ell^{\prime}} \rangle -
    N_{\ell^{\prime}} )}{B_{\ell^{\prime}}^{2}},
\end{equation}
where \mkk\ is the mode-mode coupling matrix, $N_{\ell}$ is the noise
bias, and $B_{\ell}$ is the beam transfer function.  The details of
the implementation of this algorithm are identical to the analysis of
CIBER data which uses the same detector, and can be found in the
Supplementary Materials to \citet{Zemcov2014}.

\section{Results}
\label{S:results}

These data allow us to determine both the brightness of the atmosphere
and the temporal stability of the atmospheric emission from $\sim 10$
to $\sim 100 \,$ s time scales.  

\subsection{Sky Brightness and Image Noise}

The per-integration mean sky brightness shown in Figure
\ref{fig:skybrightness} is computed by calculating the mean image
brightness in the unmasked regions of the calibrated images.  Since
stars with $J > 16$ are masked, the brightness is dominated by
atmospheric emission with a brightness of $\sim 5000 \,$\nw\, and a
small contribution from Zodiacal light (ZL) with a brightness of $\sim
300 \,$\nw \citep{Kelsall1998}.  The mean sky brightness over the
entire stable data period is $9.57 \times 10^{-2} \,$\eps, which
corresponds to $5330 \,$\nw, as shown in Figure
\ref{fig:skybrightness}.  The root mean squared variation over this
period is $\pm 30 \,$\nw.  There is no evidence for a temporal drift
in the brightness of the images during the stable period.

Because of the \oh\ emission, it is impossible to draw meaningful
comparisons between measurements of the band-averaged $J-$band sky
brightness (see e.g.~\citealt{Leinert1998}) and our measurement of the
inter-line sky brightness.  However, the \lamp\ measurement can be
compared directly with the result of \citet{Sullivan2012} who
determined the inter-line continuum between $0.83$ and $2.5 \, \mu$m
at $R=6000$ using Magellan/FIRE from Las Campanas in Chile.  At the
\lamp\ operating wavelength of $1191.3 \,$nm, they find $\lambda
I_{\lambda} = 19.8 \pm 0.15$ \magas.  They further find mean
sky brightnesses in $Y-$ and $J-$band of $20.05 \pm 0.04$ and $19.55
\pm 0.03 \,$\magas\, respectively, which at $1191.3 \,$nm
interpolates to $5287 \pm 128 \,$\nw, in good agreement with our
determination.  This is evidence that, once time-dependent emission
has quieted, the inter-line continuum at this wavelength is not
dominated by terrestrial light pollution \citep{Leinert1998}.
\citet{Sullivan2012} show that their measurements are consistent with
previous measurements at $1.6 \, \mu$m where the inter-line background
is up to $0.7$ \magas\ brighter (\citealt{Maihara1993},
\citealt{Cuby2000}, \citealt{Ellis2012}).

We can also use the Gemini Observatory sky
model\footnote{\url{http://www.gemini.edu/sciops/telescopes-and-sites/observing-condition-constraints/ir-background-spectra\#Near-IR-short}}
to compute the atmospheric emission spectrum.  The model is computed
with the sky transmission files generated by ATRAN \citep{Lord1992}
scaled to a $273 \,$K blackbody.  We assume $5 \,$mm of precipitable
water vapor during our measurements, which is consistent with the
typical value above Table Mountain during clear weather
\citep{Leblanc2011}, and find a prediction of $19.5\,$\magas\ from the
model.  Subtracting the estimate for ZL emission, this is $\sim
25\,$\% greater than our measurement, which also agrees with the
conclusion of \citet{Sullivan2012} that the Gemini model over-predicts
the inter-line sky brightness at $J-$band by a similar factor.

In addition to the mean brightness, it is useful to calculate the
image noise properties to show they track the simple theoretical
predictions discussed in Section \ref{sS:theorysens}.  This allows
us to diagnose whether there is some gross source of excess noise in
the system.  The theoretical noise properties are calculated from the
quadrature sum of the read and photon contributions given by Equations
\ref{eq:readnoise} and \ref{eq:photonnoise}.  Table
\ref{tab:noisevals} summarizes the theoretical noise characteristics
of these \lamp\ measurements given the parameters listed in Table
\ref{tab:instparams}.  From the ratio $\sigma_{\rm read} / \sigma_{\rm
  photon}$ we conclude that these $104 \,$s integrations are read
noise dominated by a factor of 1.3.  At the measured sky brightness,
the integrations should become photon noise dominated after 136s of
integration.

\begin{table}[ht]
\centering
\caption{\lamp\ Characteristic Noise Properties. \label{tab:noisevals}}
\begin{tabular}{lc}
\hline
$\sigma_{\rm read}$ & $4.4 \times 10^{-2} \,$\eps \\
Sky $\langle \lambda I_{\lambda} \rangle$ & $0.96 \times 10^{-2} \,$\eps \\
$\sigma_{\rm photon}$ & $3.3 \times 10^{-2} \,$\eps \\
$\sigma_{\rm read} / \sigma_{\rm photon}$ & 1.3 \\ 
$\sigma_{\rm total}$ & $5.5 \times 10^{-2} \,$\eps \\ \hline
Theoretical $\sigma_{\rm total}$ & $3.05 \times 10^{3} \,$\nw \\ 
Measured $\sigma_{\rm total}$ & $3.07 \pm 0.08 \times 10^{3} \,$\nw \\  \hline
\end{tabular}
\end{table}

\begin{figure}[p]
\centering
   \resizebox{\hsize}{!}{\includegraphics{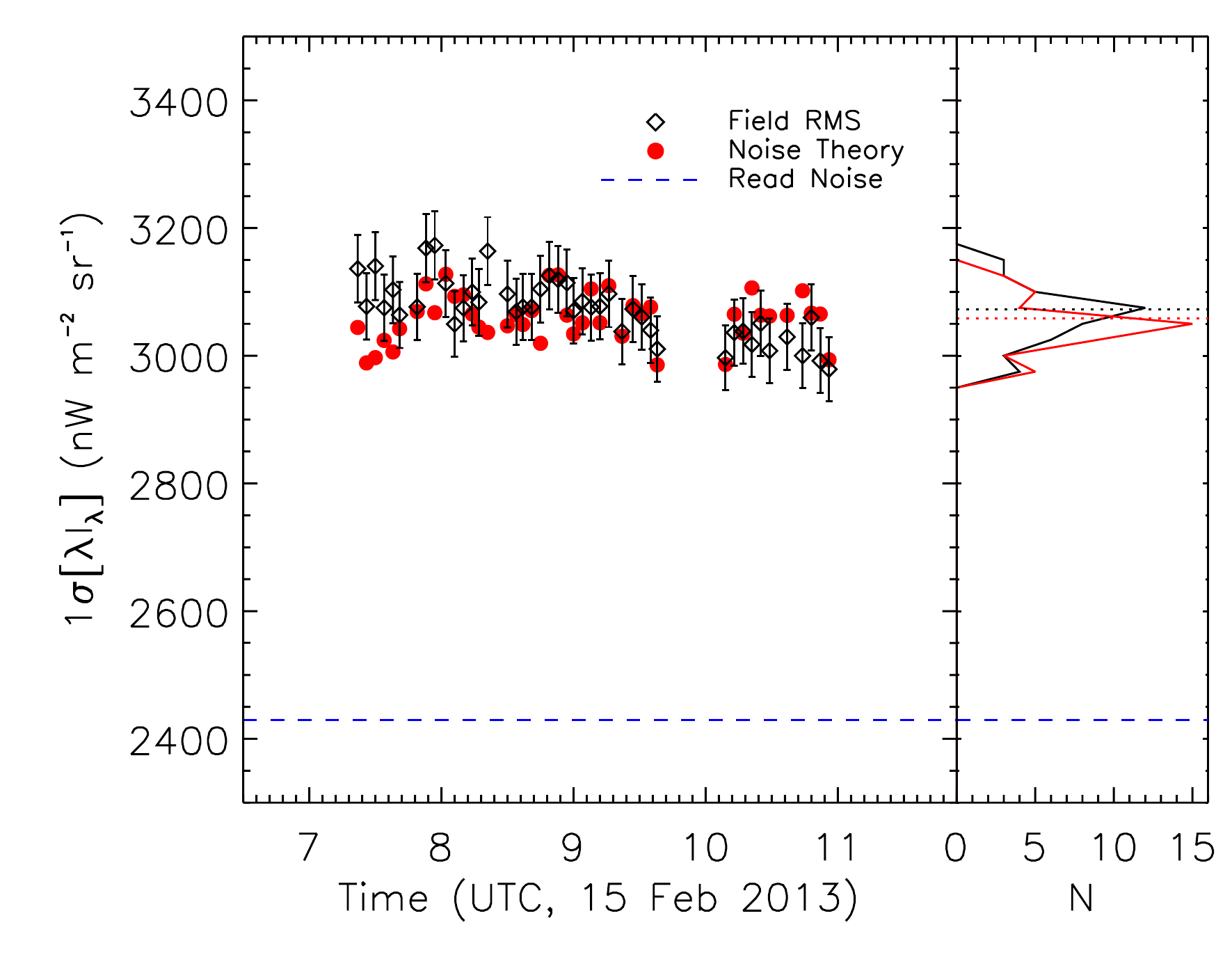}}
   \caption{Image-space noise over time.  The image space noise is
     estimated by computing the standard deviation of unmasked pixels
     in pair-wise differences of the stable data integrations
     (diamonds).  The plotted uncertainties are the errors on the
     standard deviation.  Also shown are the predictions from noise
     theory for the noise in each measurement (circles) calculated
     with the mean surface brightness of the field pairs to calculate
     the shot noise term.  The read noise is modelled as static over
     time (dashed line).  There is a weak downward trend in the noise
     properties of the images over time, but it is not clear whether
     this is due to some property of the instrument, to slight changes
     in the mean surface brightness of the sky over time, or random
     statistics.  In the right hand panel we show histograms of the
     points (solid lines) with the mean of both distributions
     indicated (dotted lines).  The $\chi^{2}$ of (data - theory) is
     $48.8$ for $44$ degrees of freedom, giving a probability to
     exceed $\chi^{2}$ of $0.29$.  We conclude that there is no
     evidence for an extra noise component in the image
     data. \label{fig:skynoise}}
\end{figure}

To compare the noise properties of the data images to the predictions,
we compute the standard deviation of unmasked pixels in time-wise
full-integration pair differences.  We remove variance from faint
sources of astronomical emission by aligning the two input
integrations with each other using the astrometry solution, leaving a
set of 45 differences in the data set.  The pixel distribution in the
differenced images matches a Gaussian distribution whose width is
$\sqrt{2}$ larger than $\sigma_{\rm read}$ listed in Table
\ref{tab:noisevals}.  The variance in the width of the noise
distribution is given by $S_{\rm error}^2= \sigma_{\rm total}^2
\sqrt{2/(n-1)}$ where $n$ is the number of unmasked pixels and $S_{\rm
  error}^{2}$ is the variance in the standard deviation.  These are
shown in Figure \ref{fig:skynoise} for each pair-wise integration
difference in the stable period data set.

The mean uncertainty over the stable period data is found by computing
the uncertainty-weighted mean of difference images, giving $3070 \pm
80 \,$\nw, where the total uncertainty is estimated from the noise
weights.  To check for consistency between the theoretical estimate
and measure noise performance, we compute the $\chi^{2}$ of the data
and theory points shown in Figure \ref{fig:skynoise} and find a
probability to exceed $\chi^{2}$ of 0.29, indicating consistency
between the estimates even assuming simple the uncorrelated read noise
at the level shown in Table \ref{tab:noisevals}.

\subsubsection{Sub-integration Image Noise Properties}

An important component of this study is an investigation of the noise
behavior over intervals shorter than a full $104 \,$s integration.
To do this, we calculate estimates for the photocurrent for fractional
parts of the full integration length, subtract adjacent fractional
integrations, and calculate the noise in the same manner as for the
full integrations.  The variation of the mean values are calculated in
the same way as for the full integration differences.  Because the
full integration time is $\sim 100 \,$s, we have two possible
populations of such differences: (i) differences of sub-integrations
in the same integration, limited to $T_{\rm int} < 51 \,$s, and (ii)
differences of sub-integrations in neighboring integration sets, which
allow us to add points $52 \leq T_{\rm int} \leq 104 \,$s.
Measurements of the latter type are actually separated by $> 104 \,$s
rather than being truly temporally contiguous, but they allow us to
map out the reduction in noise with integration time which should
follow the quadrature sum of Equations \ref{eq:readnoise} and
\ref{eq:photonnoise}.

\begin{figure}[ht]
\centering
\resizebox{\hsize}{!}{\includegraphics{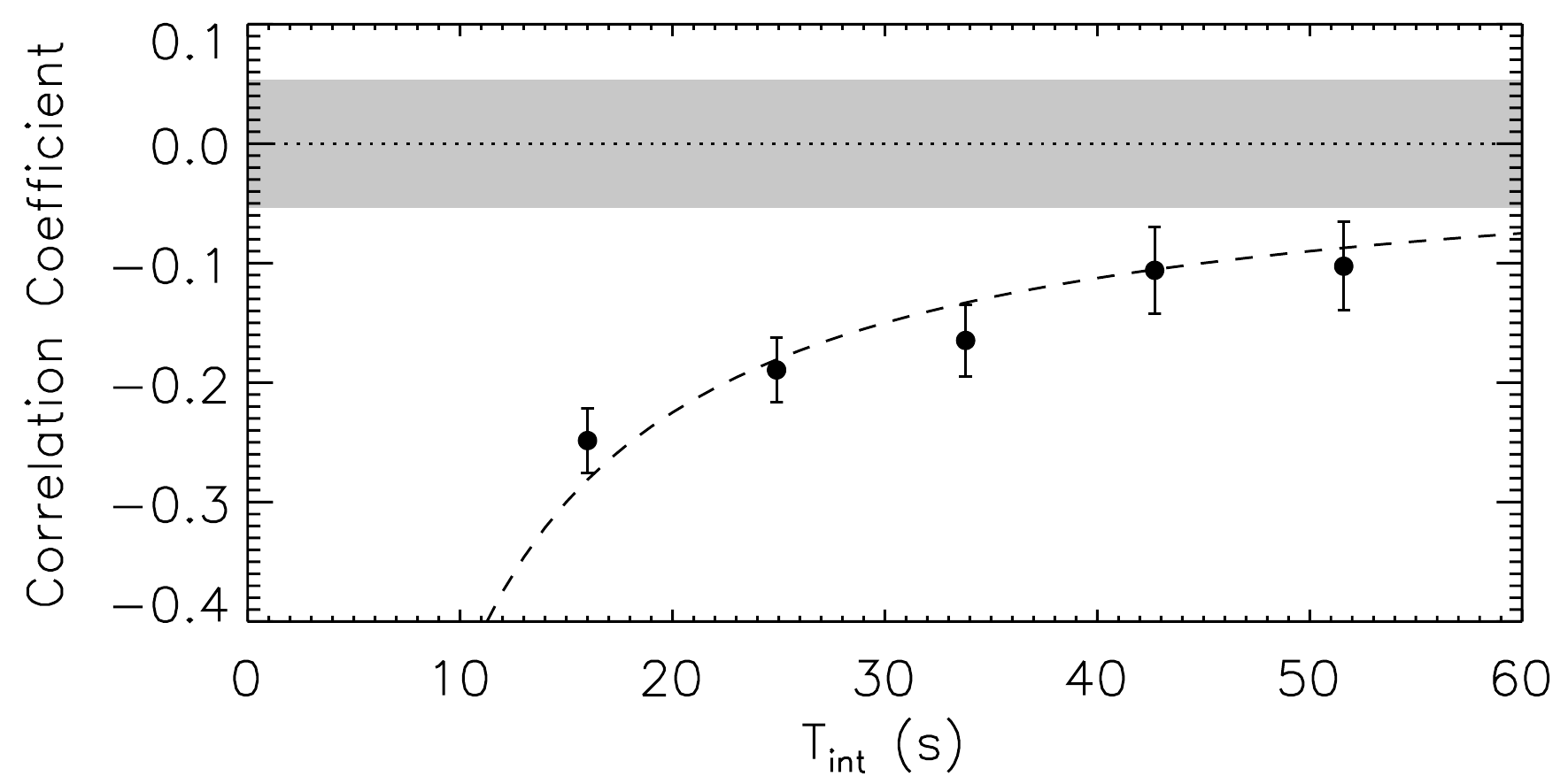}}
\caption{Correlation coefficient of sub-integration fits drawn from
  the same integration.  Because each read in a pixel of a charge integrating
  detector is correlated with the previous read, sub-integrations are
  mildly correlated with one another.  We empirically determine the
  correlation coefficient $r$ between the sub-integrations sharing a full
  integration, and plot the mean and standard deviation versus
  $T_{\rm int}$.  The scaling is well explained by $1/T_{\rm int}$
  (dashed line).  The correlation coefficient of neighboring full
  integrations are not correlated, and have a standard deviation
  about zero shown as the grey band.   \label{fig:noisecorr}}
\end{figure}

Correlated noise could modify the relationship between difference pair
noise RMS and $T_{\mathrm{int}}$ at short integrations.  As discussed
in \citet{Bock2013} and \citet{Zemcov2014}, the HAWAII-1 detectors
used in \lamp\ have correlated noise on the output with correlation
lengths of several seconds, so differencing on short time scales may
actually reduce the measured noise figure.  To check for this, we
compute the cross-correlation of the images from adjacent integrations
for sub-integrations of the first type (for $T_{\rm int} < 51 \,$s).
The resulting correlation coefficients are shown in Figure
\ref{fig:noisecorr}.

The short $T_{\mathrm{int}}$ difference images are correlated, as we
might expect for amplifier noise.  We do not measure a correlation in
the sub-integration differences taken with neighboring data sets.  To
correct the measured noise RMS for this correlation, we compute:
\begin{equation}
\label{eq:noisecorr}
\sigma^{\prime}_{i-j} = \frac{\sigma_{i-j}}{1-r}
\end{equation}
where $\sigma_{i-j}$ is the measured RMS in the difference between
sub-integrations $i$ and $j$, $r$ is the measured correlation between
them, and $\sigma^{\prime}_{i-j}$ is the corrected pixel RMS.  There
is no evidence for correlations between differences involving
sub-integrations drawn from separate major integration intervals.

\begin{figure}[htp]
\centering
   \resizebox{\hsize}{!}{\includegraphics{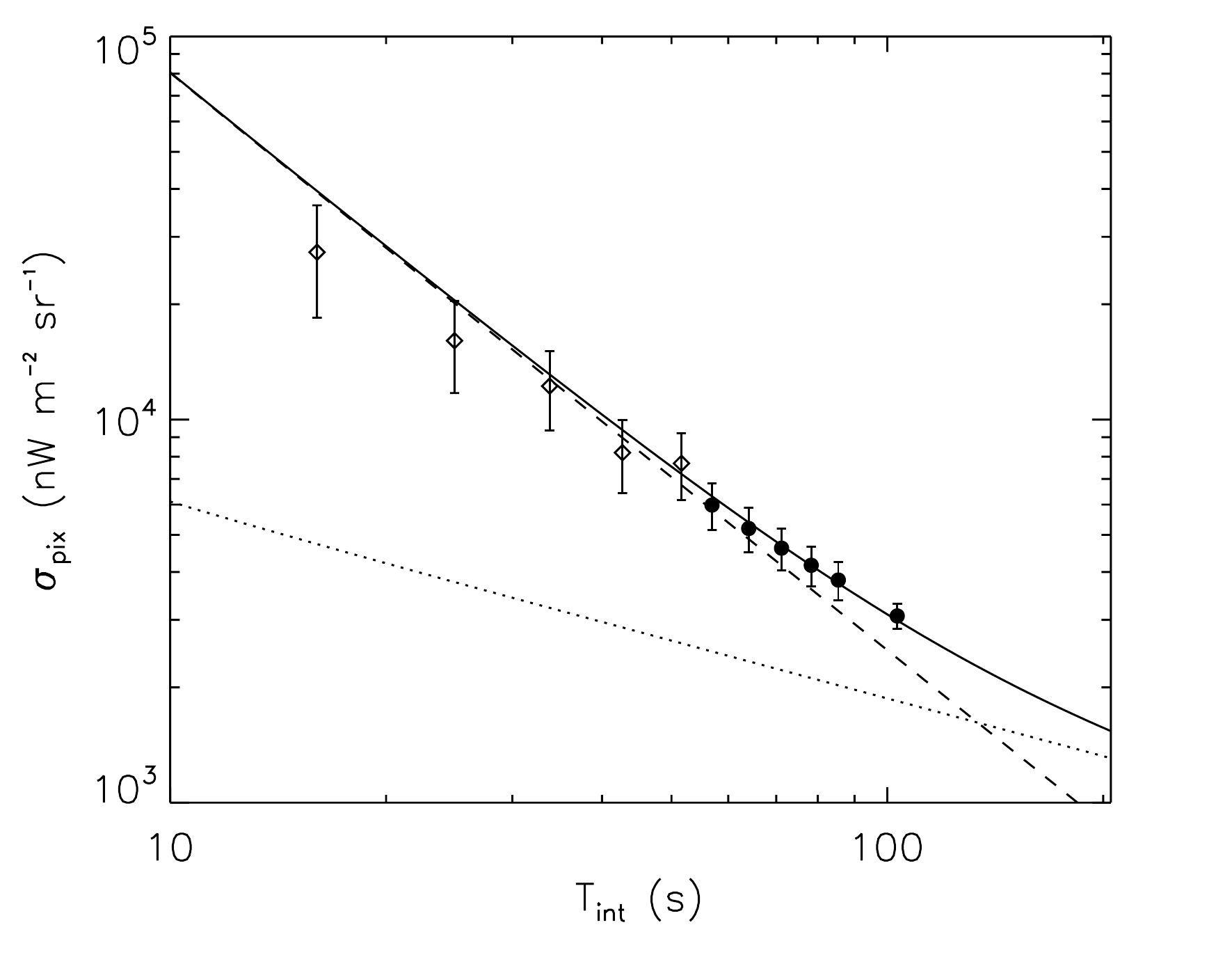}}
   \caption{The pixel standard deviation in differences of neighboring
     integrations (solid circles) or sub-integrations sharing a full
     integration (open circles) versus integration time.  The
     sub-integration points have been corrected for the effect of
     noise correlations using the empirical scaling shown in Figure
     \ref{fig:noisecorr}.  In the absence of extra noise components,
     the points should follow the sum (solid line) of shot noise
     (dotted line) and read noise (dashed line).  Known noise
     components explain the behavior of the image space pixel noise as
     a function of integration time.  \label{fig:noisermsvst}}
\end{figure}

The pixel RMS in both types of sub-integration measurements are shown
in Figure \ref{fig:noisermsvst} as a function of the effective
integration time.  For the first-type sub-integration difference RMS,
we correct for the effects of the correlation, and plot
$\sigma^{\prime}_{i-j}$.  The model effectively predicts the behavior
of the noise RMS within the uncertainties.

\subsection{Spatial Fluctuations}

We measure the time variation in spatial fluctuations in the sky
emission using auto-power spectrum measurements.  To constrain a
hypothetical component associated with variations in the continuum
level between the Meinel bands, it is necessary to account for the
other sources of power in the power spectrum, namely: (i) astronomical
emission $\widetilde{C_{\ell}}$; (ii) noise bais from the detector
$N_{\ell}^{\rm read}$; and (iii) noise bias from photon noise
$N_{\ell}^{\rm photon}$.  To account for these, we can begin by
writing equation \ref{eq:ps} in a slightly different form in which we
solve for the measured sky and ignore the effects of image masking:
\begin{equation}
\label{eq:psinv}
\langle  C_{\ell} \rangle = B_{\ell}^{2} \widetilde{C_{\ell}} + N_{\ell},
\end{equation}
where $N_{\ell} = N_{\ell}^{\rm read} + N_{\ell}^{\rm photon} +
N_{\ell}^{\rm sky}$, where $N_{\ell}^{\rm sky}$ is the hypothetical
component of the noise due to time variations in the spatial emission
of the atmosphere.

Ideally, to isolate $N_{\ell}^{\rm sky}$, we would difference away the
other terms in the sum in Equation \ref{eq:psinv}.  The
$\widetilde{C_{\ell}}$ term is common to neighboring integrations, and
so can be cancelled by differencing neighboring integrations which
have been aligned to one another.  However, because the noise is
different in each realization of the measurement it is not possible to
cancel the $N_{\ell}$ terms through differencing of data alone.
Rather, to isolate this component, we observe that from Equations
\ref{eq:readnoise} and \ref{eq:photonnoise} the time-dependence of the
components $N_{\ell}^{\rm read}$ and $N_{\ell}^{\rm photon}$ can be
inferred.  That is, by differencing integrations aligned to one
another $\widetilde{C_{\ell,1}} = \widetilde{C_{\ell,2}}$, so we have:
\begin{align}
\label{eq:noiseps} 
\langle  C_{\ell}^{\rm diff} \rangle & =  N_{\ell,1}^{\rm read} + N_{\ell,1}^{\rm photon} +
N_{\ell,1}^{\rm sky} + N_{\ell,2}^{\rm read} + N_{\ell,2}^{\rm photon} +
N_{\ell,2}^{\rm sky} \\
& =  (N_{\ell,1}^{\rm read} + N_{\ell,2}^{\rm read}) (T/T_{0})^{-3} + 
(N_{\ell,1}^{\rm photon}  + N_{\ell,2}^{\rm photon}) (T/T_{0})^{-1} + 
(N_{\ell,1}^{\rm sky}  + N_{\ell,2}^{\rm sky}), \notag
\end{align}
this is equivalent to:
\begin{equation}
\label{eq:noisetimeps} 
\langle  C_{\ell}^{\rm diff}(t) \rangle  =  a_{\rm read}T^{-3} + a_{\rm photon}T^{-1} + a_{\rm off},
\end{equation}
where $\langle C_{\ell}^{\rm diff} \rangle$ is the difference power
spectrum and the $a_{i}$ are constants of proportionality related to
$N_{\ell}$.  The scheme we use here is to measure $\langle
C_{\ell}^{\rm diff} \rangle$ as a function of time and to constrain a
component which does not behave as either of the known noise
components.  We note that by using this model we are constraining a
time-invariant term in the noise $a_{\rm off}$ which we correspond to
the $N^{\rm sky}$ term, and that we would not be at all sensitive to
an excess component which behaves as either $T^{-3}$ or $T^{-1}$.
Since this is a preliminary measurement, and we expect any deviation
from the expected behavior of the noise to change slowly with $T_{\rm
  int}$, constraining the simplest possible model (\textit{i.e.} an
offset in the model) is a conservative approach.

\begin{figure}[htp]
\centering
\resizebox{\hsize}{!}{\includegraphics{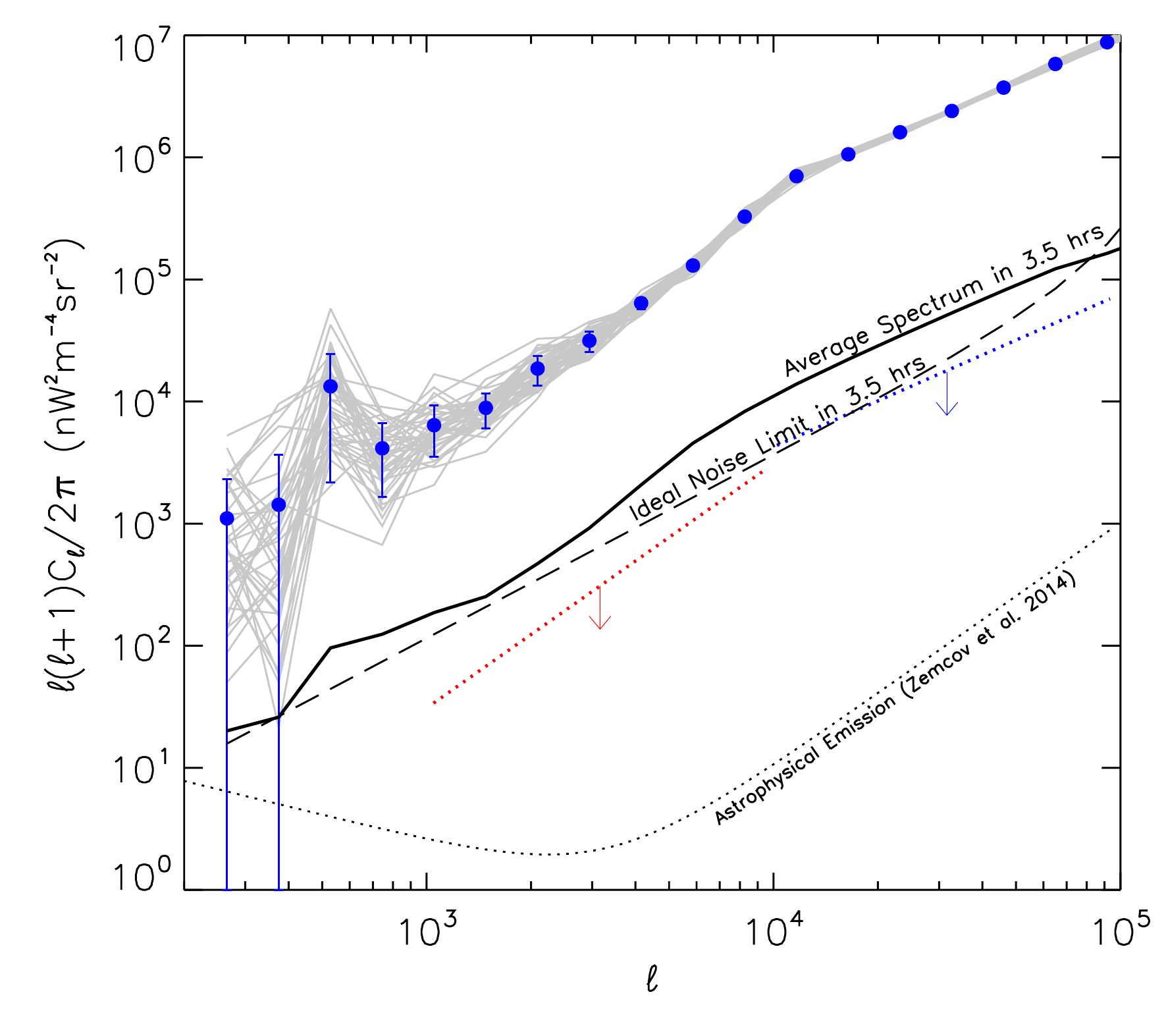}}
\caption{\lamp\ power spectra for full integration time field
  differences.  The grey lines show the individual power spectra for
  each of the 45 differences.  The blue points show the mean power
  spectrum, with uncertainties plotted encompassing the per bandpower
  standard deviation over the ensemble.  The solid black line shows
  the total sensitivity reached by averaging all of the individual
  integrations together, and the dashed black line shows the
  theoretical statistical sensitivity in the full data set assuming
  uncorrelated white noise. Though there is evidence for some excess
  noise above the ideal noise limit, it is difficult to assign to
  correlated read noise or an excess noise component from this plot
  alone.  The red and blue dotted lines show $2 \sigma$ upper limits to a
  component of the noise that does not follow the time-dependence of
  either read or photon noise.  We find no evidence for an excess
  component in the noise to the limit of the measurement.  The dotted
  line shows the fiducial astrophysical power spectrum for $J > 17.5$
  source masking measured by \citet{Zemcov2014}.  The noise is larger
  than the astronomical power from IHL, but $250$ hours of integration
  time would decrease to interesting levels a factor of $\sim 70$
  lower than the currently achieved noise. \label{fig:powerspectra}}
\end{figure}

Figure \ref{fig:powerspectra} shows the full set of $\langle
C_{\ell}^{\rm diff} \rangle$ for the set of 45 full-integration time
difference field data scaled to the non-differenced amplitude.  The
predicted effective sensitivity for the combined 3.5 hrs of
integration lies above the ideal sensitivity calculated assuming
uncorrelated white noise with $\sigma = 3.07 \times 10^{3} \,$\nw.
Because these detectors have complex noise properties
\citep{Zemcov2014}, the simple comparison is not diagnostic as the
excess noise component may be due to read noise correlations, or a
component due to sky variability.

\begin{figure}[htp]
\centering
\resizebox{\hsize}{!}{\includegraphics{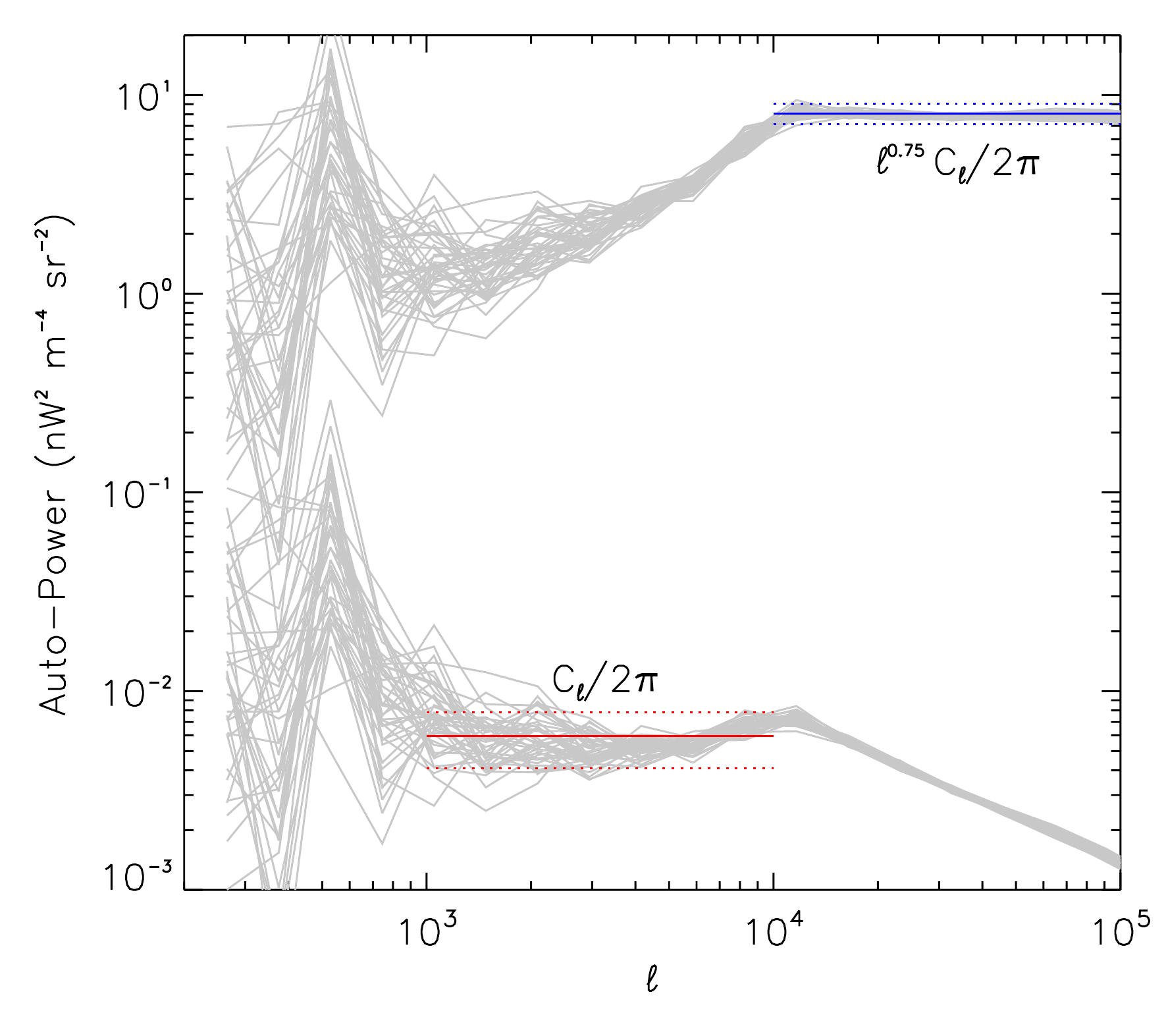}}
\caption{The \lamp\ power spectra show in Figure
  \ref{fig:powerspectra} using different scalings.  The pre-factors,
  $1/2\pi$ in the $10^{3} < \ell < 10^{4}$ region and $\ell^{0.75}/2
  \pi$ in the $10^{4} < \ell < 10^{5}$ region, are chosen to make the
  power approximately constant in the regions of interest.  We ignore
  the $\ell < 10^{3}$ region due to large variance.  The mean and
  standard deviation of the set of 45 measurements in the two regions
  are indicated by the solid and dashed lines.  These statistics are
  calculated for the full set of $T_{\rm int}$ measurements to develop
  the behavior of the noise with integration
  time. \label{fig:psfunny}}
\end{figure}

To succinctly compress the power information in the power spectra, for
each $T_{\rm int}$ we compute the mean of all bandpowers in two broad
bins, one between $10^{3} < \ell < 10^{4}$ and the other $10^{4} <
\ell < 10^{5}$, as shown in Figure \ref{fig:psfunny}.  As these two
regions are made flat with different $\ell$ scalings, we chose a
multiplicative prefactor of $1/2\pi$ in the lower-$\ell$ region and
$\ell^{0.75}/2\pi$ in the higher-$\ell$ region.  This allows us to
compute a mean over a region with approximately the same value.

\begin{figure}[htp]
\centering
\resizebox{\hsize}{!}{\includegraphics{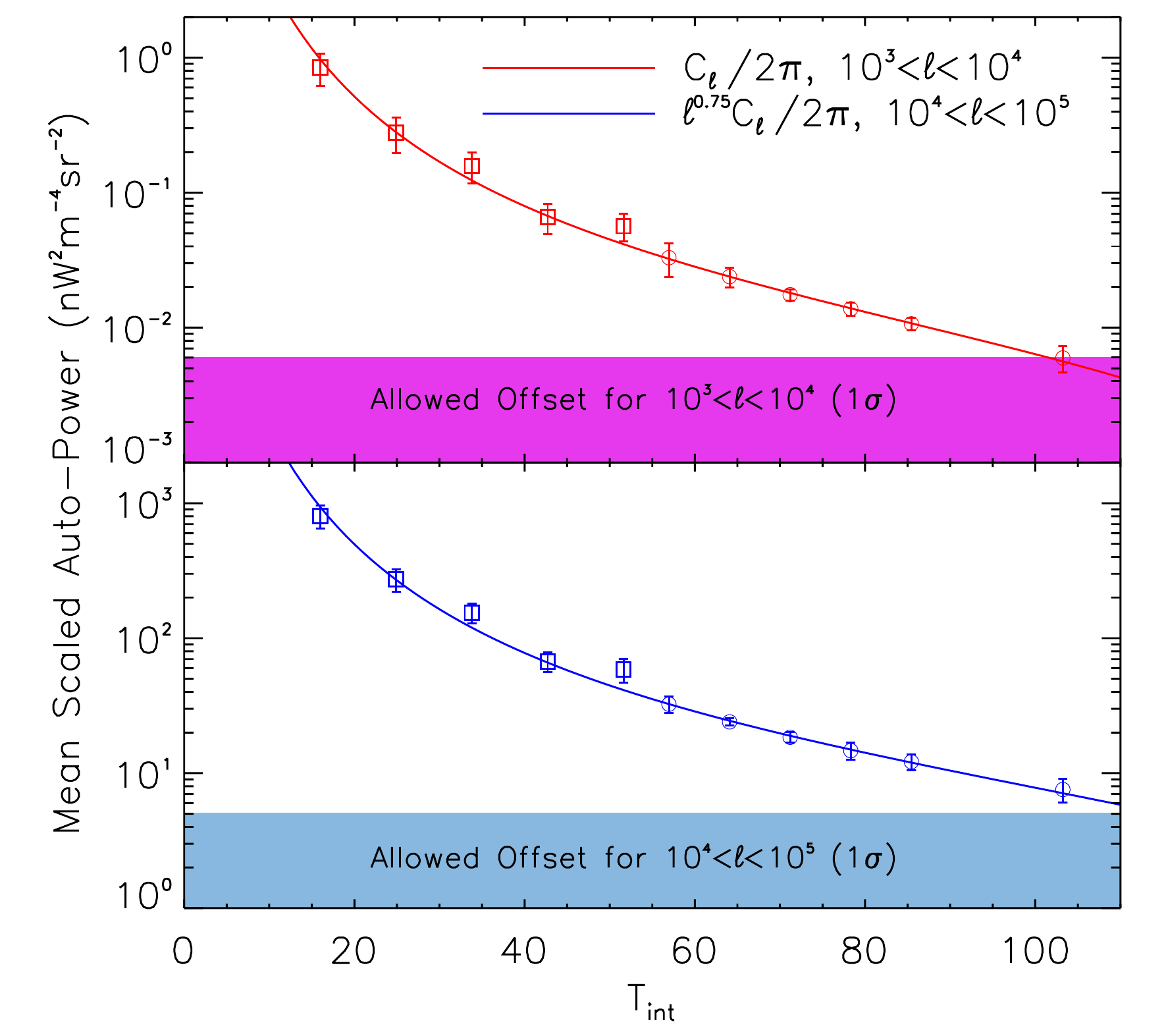}}
\caption{Scaled noise power versus integration time for the \lamp\
  field-difference data.  The upper plot shows the mean power at each
  value of $T_{\rm int}$ for the $10^{3} < \ell < 10^{4}$ bandpower
  averages, and the lower panel the same for $10^{4} < \ell < 10^{5}$.
  The lines indicate the best fitting model of the form $C_{\ell} =
  a_{\rm read} T^{-3} + a_{\rm photon} T^{-1} + a_{\rm sky}$ in both
  cases.  Finally, the colored regions show the allowed $1 \sigma$
  uncertainty $a_{\rm off}$ from the fit, which we use to place
  constraints on the amplitude of noise which arises from a
  hypothetical component associate with emission from the sky.
  \label{fig:psvstint}}
\end{figure}

The time-dependence of the bandpower averages is shown in Figure
\ref{fig:psvstint}, which show the mean of the two $\ell$ regions
versus $T_{\rm int}$ for all differencing time scales we probe in this
measurement.  The $T_{\rm int} < 55 \,$s difference sets, which are
computed from the same integrations, are corrected by a factor of
$r^{2}$ to account for their correlation.  To constrain the value of
$a_{\rm sky}$ in Equation \ref{eq:noisetimeps}, we fit a function of
that form to the scaled mean as a function of $T_{\rm int}$.  The fit
results in an offset consistent with zero within uncertainties, which
is good evidence there is not a component of the noise which does not
scale with integration time.  To place limits on the amplitude of this
component in the power spectra, we compute the $2 \sigma$ upper limit
from the best-fitting value of $a_{\rm off}$ and compute its ratio to
the $T_{\rm int} = 104 \,$s point in Figure \ref{fig:psvstint}.  We
then scale the amplitude of the power spectra shown in Figure
\ref{fig:powerspectra} at the mean bandpower by this ratio to generate
an upper limit on the atmospheric variation as a function of $\ell$,
as shown in Figure \ref{fig:powerspectra}.  We do not detect an excess
noise component which does not scale with time in these data.

\subsection{Systematic Uncertainties}
\label{sS:systematics}

The major systematic uncertainties which affect this work are the dark
current and flat field corrections.  To place limits on these, we
compute the power spectra in the following cases: (i) when no flat
field is applied; (ii) when the dark current correction is varied by
its uncertainty in the positive direction; and (iii) when the dark
current correction is varied by its uncertainly in the negative
direction.

We motivate the flat field systematic check by noting that the
photocurrent at the detector is very small, so the gain errors have a
very small effect compared to the variance of the measurement.  In
measurements of the mean or variance of images, the flat field
correction will have very little effect since it is referenced to the
mean of the measurements.  The flat field will have more of an effect
in power spectral measurements, as it has spatial coherence over large
scales in the array \citep{Bock2013}.  As a result, we expect that the
flat field will have little effect except at low $\ell$ modes.

Because of the low photocurrent in these measurements, the dark
current correction is the single most important source of systematic
uncertainty.  The dark current estimate, which is formed per pixel
from the ensemble variance of the dark current measurements, has small
uncertainties, but both the correction and its uncertainties do
exhibit large scale structure, so we are most concerned about the
presence of systematic uncertainties in the spatial power spectra.  We
place limits on the effect of the dark current correction by computing
power spectra in which the dark current correction is varied by its $1
\sigma$ uncertainty in both the positive and negative directions.
This simulates the effect of over- and under-estimating the dark
current correction, respectively, and is a conservative upper limit to
the size of the possible coherent effect we would expect.

\begin{figure}[p]
\centering
\resizebox{\hsize}{!}{\includegraphics{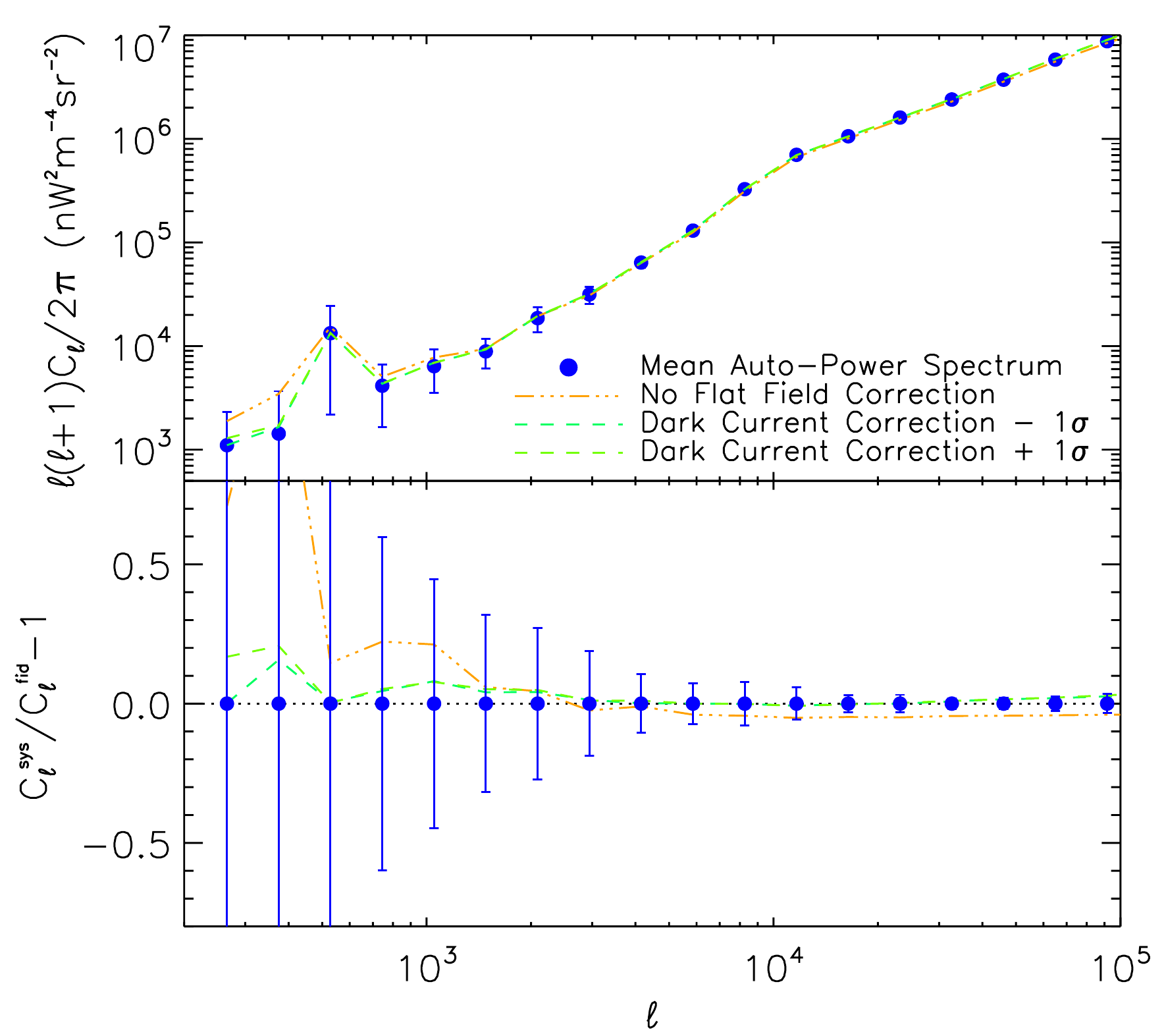}}
\caption{Sources of systematic errors and limits on their effect on
  the mean auto-power spectrum.  To investigate the effect of the two
  major systematic uncertainties in this study, namely the flat field
  responsivity correction and the dark current subtraction, we
  compute the auto-power spectra in the fiducial case (points), and
  then (i) with no flat field correction applied; (ii) with the dark
  current subtraction under-estimated by the overall variance in the
  set; and (iii) with the dark current subtraction over-estimated by
  the overall variance in the set.  The upper panel shows power
  spectra scaled by $\ell (\ell+1) / 2 \pi$ which can be compared directly
  to Figure \ref{fig:powerspectra}.  The lower panel shows the
  ratio between the fiducial power spectrum and the systematic test of
interest.  The flat field systematic has the largest effect on the
measurement, overestimating the power at low-$\ell$ and slightly
underestimating the power at high-$\ell$.  When propagated to overall
uncertainty on the atmospheric noise component, we find at most a $5
\,$\% variation from the fiducial upper limit.}
\label{fig:pssystematics}
\end{figure}

Figure \ref{fig:pssystematics} shows the systematic uncertainties
arising in the power spectra from these tests.  The systematic
uncertainties are typically manageable for $\ell < 10^{3}$ where they
are $< 20 \,$\% at all bandpowers, and $< 5 \,$\% for $\ell < 3000$.
As expected, the flat field correction is the largest source of
systematic uncertainty, showing systematic deviation above the
fiducial power spectrum for $\ell < 3000$, and below it for $\ell >
3000$.  We propagate these power spectra systematic errors to upper
limits on the atmospheric noise component shown in Figure
\ref{fig:powerspectra} and summarize the results in Table
\ref{tab:systematics}.  We quote $\ell (\ell+1) \langle \delta
C_{\ell, \mathrm{sky}}\rangle / 2 \pi$, which here we define to be the
$2 \sigma$ upper limit on auto-power over the two $\ell$ ranges
equivalent to the upper limits plotted in Figure \ref{fig:powerspectra}. As
these reflect systematic uncertainties due to various corrections in
the data pipeline we cannot add them to the overall uncertainty in the
limit, but based on the amplitude of the changes in the noise limits
we can infer that these systematic errors make at most a $5 \,$\%
change in the upper limit we quote.  Similar analyses with larger
data sets will require care with these type of systematics, but they
do not limit our understanding of the data in this study.

\begin{table}[ht]
\centering
\small
\caption{Systematic uncertainties and their effect on power spectral noise
  upper limits. \label{tab:systematics}}
\begin{tabular}{lccc} 
\hline
Systematic & Estimate & \multicolumn{2}{c}{$\ell (\ell+1) \langle \delta C_{\ell,
  \mathrm{sky}}\rangle / 2 \pi$ (\nw)} \\ 
Error & & $10^{3} < \ell < 10^{4}$ $(\times 10^{-2})$ & $10^{4} < \ell <
10^{5}$ $(\times 10^{-4})$ \\ \hline
Fiducial & - & $3.09$ & $1.74$ \\
Flat Field & No flat field correction & $3.02$ & $1.83$ \\
Dark current & Dark current correction $-1 \sigma$ & $3.02$ & $1.73$ \\
$^{\prime \prime}$ & Dark current correction $+1 \sigma$ & $3.01$ & $1.73$ \\ \hline
\end{tabular}
\end{table}

\section{Discussion}
\label{S:discussion}

To the limit we are able to probe with these data, variations in the
inter-line continuum level at \lamplambda\ do not appear to be causing
a deviation from normal integration-time noise scalings for this
detector.  Though the total integration time on the sky of $3.5 \,$hrs
is not sufficient to approach the diffuse astronomical emission, this
result lends confidence to the plausibility of performing narrow-band
imaging measurements in the near IR.  \citet{Sullivan2012} investigate
the possible sources of the terrestrial continuum background near $1.2
\, \mu$m, and are unable to identify the source of the inter-line
continuum, finding it to be larger than the sum of known contributions
at these (namely the Lorentzian wings of the \oh\ lines and Zodiacal
light).  Our results, from an observation site and instrument with
significantly different characteristics, are remarkably consistent
with those of \citet{Sullivan2012} and suggest that their measurement
of the continuum level is correct.  Because the source of the
inter-line continuum remains a mystery, it is dangerous to speculate
whether these observations would eventually become limited by the
background continuum.  A lower limit to the temporal variability can
be set using the model of \citet{Sullivan2012} which gives an \oh-wing
background of $\sim 500 \,$\nw\ at \lamplambda.  Assuming $10 \,$\%
variation in amplitude, the expected noise level from the \oh-wings is
$\sim 1.5$ orders of magnitude below our current sensitivity, so it is
not clear whether this component would present a fundamental limit to
measurements of the near IR background.

Another open issue is the time scale of the emission, which is known
to vary on the time scale of minutes (\citealt{Taylor1991},
\citealt{Ramsay1992}).  Broad-band measurements of airglow in $H-$band
show a temporal powers spectrum rising as $1/f$ with a knee at $\sim 5
\,$mHz\footnote{Measured by the 2MASS Wide-Field Airglow Experiment,
  \url{http://www.astro.virginia.edu/~mfs4n/2mass/airglow/adams/syp.ps}.}.
Because our spatial power spectra are differences of $\sim 100 \,$s
integrations, we are probing the airglow stability on timescales
similar to this.  Unlike high-spectral resolution instruments
\citep{Sullivan2012}, we are not concerned about time scales longer
than a few minutes because it is possible to design an $R \sim 500$
narrow-band imaging instrument to be photon-noise limited in $\sim 100
\,$s. 

Though based on a limited data set, these results are encouraging for
the general approach of imaging the near IR background through narrow
windows in the Meinel emission.  With a suitable instrument,
observation design, and careful attention to systematics, it seems
realistic to achieve the sensitivity required to measure diffuse
astronomical emission from the ground.

\begin{acknowledgements}
  The authors wish to thank Jaime Luna for his help designing the \lamp\
  mechanical assembly, Heath Rhoades at JPL's Table Mountain
  Observatory for his assistance setting up the instrument and
  guidance using the $24^{\prime \prime}$ telescope, and the Gemini
  Observatory for making their sky model tables public. The
  development of \lamp\ was supported by the JPL Research and
  Technology Development Fund.  This publication makes use of data
  products from the Two Micron All Sky Survey (2MASS), which is a
  joint project of the University of Massachusetts and the Infrared
  Processing and Analysis Center/California Institute of Technology,
  funded by the National Aeronautics and Space Administration and the
  National Science Foundation.
\end{acknowledgements}

\bibliographystyle{apj}                       %% AASTeX

\end{document}